\documentclass[showpacs,preprint,amsmath,amssymb,floatfix]{revtex4}

\usepackage[dvips]{graphicx}   
\usepackage{dcolumn}           
\usepackage{bm}                
\usepackage{subfigure}
\usepackage[normalem]{ulem}              
\usepackage{color}              

\newcommand{\RED} {black}    
\newcommand{\BLUE}{black}    

\begin{document}
\title{Polyelectrolytes in Multivalent Salt Solutions under the Action of DC Electric Fields} 
\author{Kun-Mao Wu}
\author{Yu-Fu Wei}
\author{Pai-Yi Hsiao}
\email[Corresponding author. Email:\ ]{pyhsiao@ess.nthu.edu.tw} 
\affiliation{Department of Engineering and System Science, 
National Tsing Hua University, Hsinchu, Taiwan 300, R.O.C.}

\date{\today}

\begin{abstract}
We study conformational and electrophoretic properties of polyelectrolytes
(PEs) in tetravalent salt solutions under the action of electric fields by
means of molecular dynamics simulations. \textcolor{\BLUE}{Chain conformations
are found to have a sensitive dependence on salt concentration $C_s$. As $C_s$
is increased, the chains first shrink to a globular structure and subsequently
reexpand above a critical concentration $C_s^*$.  An external electric field
can further alter the chain conformation.  If the field strength $E$ is larger
than a critical value $E^*$, the chains are elongated. $E^*$ is shown to be a
function of $C_s$ by using two estimators $E_{I}^*$ and $E_{II}^*$ through the study
of the polarization energy and the onset point of chain unfolding,
respectively.  The electrophoretic mobility of the chains depends strongly on
$C_s$, and the magnitude increases significantly, accompanying the chain
unfolding, when $E> E_{II}^*$. We study the condensed ion distributions
modified by electric fields and discuss the connection of the modification with
the change of chain morphology and mobility. Finally, $E^*$ is studied by
varying the chain length $N$. The inflection point is used as a
third estimator $E_{III}^*$. $E_{III}^*$ scales as $N^{-0.63(4)}$ and
$N^{-0.76(2)}$ at $C_s=0.0$ and $C_s^*$, respectively.  $E_{II}^*$ follows a
similar scaling law to $E_{III}^*$ but a crossover appears at $C_s=C_s^*$ when
$N$ is small. The $E_{I}^*$ estimator fails to predict the critical field,
which is due to oversimplifying the critical polarization energy to the thermal
energy. Our results provide valuable information to understand the
electrokinetics of PE solutions at the molecular level and could be helpful in
micro/nano-fluidics applications.}
\end{abstract}


\maketitle

\section{Introduction}
\textcolor{\RED}{Electric field-driven molecular analysis and sorting techniques
have been widely used in many domains of research, such as chemistry, biology,
and medical engineering. While the functionality of a bio-sensing or diagnostic
system becomes more and more complicated and specialized today, electrokinetics
remains the mechanism of choice for fluid actuation and manipulation at
micrometer or submicrometer scales through the use of electric
fields~\cite{changbook10}.  Therefore, a solid understanding of the
electrokinetic behavior of ions, molecules, and macromolecules under the
influence of electric fields is necessary to successfully integrate
electric fields in micro/nano-fluidic devices.}

Electrophoresis has been developed to separate charged macromolecules, such as
DNA molecules or proteins, based upon their molecular weight for many
years~\cite{cottet98, viovy00, mayer94}. In most situations, electrophoresis is
performed in sieving mediums such as gels. This is because, in free solutions,
DNA molecules cannot be size-separated owing to the free-draining effect: the
hydrodynamic friction and the molecular charge of a DNA molecule are both
linearly proportional to the chain length.  Hence, the electrophoretic mobility
is length independent~\cite{shendruk11}.  This phenomenon applies not only to
DNA molecules but to any charged macromolecule, or polyelectrolyte (PE). In
gels, the mechanism of electrophoretic separation is dominated by biased
reptation of PEs and so the mobility has chain-length
dependence~\cite{viovy00,slater09}. One drawback of the method is its low
efficiency. Researchers continue to search for new techniques to separate
charged molecules more efficiently~\cite{shendruk11}, particularly in
micro/nano-fluidics. 
   
Recently, Netz proposed an idea to separate PEs electrophoretically in free
solutions~\cite{netz03a, netz03b}. PEs are polarized by electric fields and, if
the electric field is strong enough, the polarization can induce chain
unfolding, which renders a drastic increase in the electrophoretic mobility of
the chains. Most importantly, the critical electric field to unfold chains is
found to depend on the length.  This dependence provides a plausible manner to
separate chains  by size in free solutions through an unfolding transition. If
the idea could be applied in micro/nano-fluidics, the efficiency to separate
DNA or other biomacromolecules by size could be improved.  In Netz'
idea-demonstrating work~\cite{netz03a, netz03b}, PEs were simply collapsed by
monovalent counterions by setting a strong Coulomb coupling parameter between
charged particles. However, this situation is not realistic because the PEs are
usually collapsed by adding condensing agents such as multivalent
salt~\cite{bloomfield96}.  Coulomb coupling is not as strong as in the
simulation work and furthermore, there are several species of counterions
present in solutions which compete with  each other to condense on the chains.
The distributions of these ions play a crucial role in determining how PEs
unfold in electric fields. \textcolor{\RED}{Simulating polarization induced
unfolding in salty solutions requires that individual ions be explicitly
modeled with interactions that realistically account for aqueous conditions.}
This is what we will focus on in this study.

When salt is added in solutions, especially multivalent salt, PEs show
complicated behavior~\cite{bloomfield97, grosberg02, quesada-perez03}.  The
addition of multivalent salt can induce collapse or aggregation of chains,
which causes phase separation. An excessive addition of multivalent salt can
even cause the separated phases to dissolve back to a homogeneous solution; the
collapsed and aggregated chains reexpand and separate from each other.  Using
these phenomena, researchers are able to control the dimension of DNA molecules
in solutions and collapse them into small, very ordered toroidal
particles~\cite{conwell03}. In addition to the size control, the presence of
multivalent counterions can also modify the charge distribution around PE
chains~\cite{wei10} and lead to a specific phenomenon called ``overcharging'',
where the condensed counterions overcompensate the charge on chain
surfaces~\cite{nguyen00, grosberg02}. The total chain charge is also
effectively altered. \textcolor{\RED}{In certain conditions, the effective
chain charge changes sign, a phenomenon called ``charge
inversion''~\cite{quesada-perez03}.} A simple way to determine the effective
charge is to study electrophoresis of PEs in weak electric
fields~\cite{hsiao08a}. Controlling the charge of PE-ion complexes is an
important issue for gene therapy because it is related to the efficiency of DNA
up-take by cells through endocytosis pathways~\cite{vijayanathan02}.
 
In strong electric fields, the behavior of PEs  becomes even more complicated
because the charged particles in the complexes respond to the field in
different ways, depending on the charge and the sign. Moreover, strong electric
fields can change the molecular conformation in ways that are difficult to
predict. The electrokinetics of the system is significantly modified. It has
been demonstrated that DNA molecules condensed by polyvalent counterions such
as spermine can be decondensed in DC electric fields if the electric field
strength exceeds a threshold $E^*$~\cite{porschke85a}.  \textcolor{\RED}{Netz
predicted that $E^*$ should scale as $N^{-3\nu/2}$ where $N$ is the chain
length and $\nu$ is the Flory exponent~\cite{netz03a,netz03b}.  Simulations did
observe scaling behavior but the scaling exponent measured did not agree with the
prediction~\cite{hsiao08b, wei09, liu10}.}  Hsiao and Wu considered a coiled
chain as an ellipsoidal object of volume $V$ and proposed a modification of the
scaling law to be $E^* \sim V^{-3/2}$~\cite{hsiao08b}. The salt valency
dependence of $E^*$ has also been explored~\cite{hsiao08b, wei09, liu10}. The
results show that the magnitude of electrophoretic mobility of chain increases
significantly above a chain unfolding transition, providing the foundation for
chain separation in free solutions.  Moreover, the chains can be unfolded to an
elongated structure, which can then be utilized in the techniques of single-DNA
molecule sequencing to increase spatial resolution of detection~\cite{chan05,
gupta08, treffer10}. 

\textcolor{\RED}{The response of PEs to alternating-current (AC) electric fields
has also been investigated recently by simulations~\cite{hsiao11, liu10}. Liu
et al.~found that chains are stretched and the sizes breath with the frequency
of applied AC field only when the field strength exceeds some critical value
and the frequency is smaller than the intrinsic relaxation frequency of the
chain~\cite{liu10}. The work by Hsiao et al.~further connected the critical AC
field strength with the DC one, and the critical AC frequency with the inverse
DC chain-fluctuation time~\cite{hsiao11}. A model, based upon Maxwell-Wagner
dielectric theory, has also been developed, which explains the critical
field-frequency correlation for chain unfolding in AC fields~\cite{hsiao11}.
Recently, the conformational transition to a stretched state in AC fields has
been experimentally demonstrated~\cite{wang10}. The chain size shows
interesting hysteretic behavior upon sweeping the AC frequency.  To understand
the behavior, it is very important to first investigate chain conformations,
ion distributions, and also the mobilities of the chains and ions in DC fields,
because DC fields can be regarded as AC fields with zero frequency.}

PE solutions involve both polymeric and electrolyte degrees of freedom, which
brings many difficulties in dealing with these systems theoretically. Molecular
dynamics simulations are a simple and economic tool, able to study systems in a
controllable way and capture detailed information at the molecular level.  Many
simulation works investigate the structure of PEs in different solution
conditions~\cite{bloomfield97, wei07, wei10, hsiao06a, hsiao06b, hsiao06c} and
the response to electric fields~\cite{mccormick07a, mccormick07b, netz03a,
netz03b, hsiao08a, hsiao08b, wei09, hsiao11}. However, detailed information
about the electrokinetics of different species of ions around the chains is
lacking. Since the conformation of chains depends strongly on the concentration
of salt in solutions, it is very important to understand the distribution of
ions around the chains and see how these ions are affected by external electric
fields.  Therefore, in this study we aim to investigate electrophoresis of
single PEs in multivalent salt solutions by means of molecular dynamics
simulations.  The rest of the paper is organized as follows.  We explain the
model and simulation setup in Sec.~II. The results and discussions are
presented in Sec.~III. The effect of tetravalent salt concentration on chain
conformation is discussed first (Sec.~III-A). We then study polarization and
determine the critical electric field at various salt concentrations
(Sec.~III-B). The electrophoretic mobility of chains and condensed ions are
investigated in Sec.~III-C. The distribution of condensed ions and the
effective charge of chains are presented in Sec.~III-D and Sec.~III-E. Finally,
the mobility dependence on chain length is studied in Sec.~III-F. We give our
conclusions in Sec.~IV.

\section{Model and simulation method} 
\label{sec:model} Our system comprises a single linear chain, modeled by a
bead-spring chain model. The chain consists of $N$ monomer beads. Each bead
carries a negative unit charge $-e$ and dissociates one monovalent cation (or
``counterion'') into the solution. The bonds connecting two adjacent monomers
are modeled by the finitely extensible nonlinear elastic (FENE) potential
\begin{equation}
U_{\rm FENE}(b)= -\frac{1}{2}k b_{\rm max}^2 \ln\left(1-\left(
\frac{b^2}{b_{\rm max}^2}\right)\right)
\end{equation}
where $b$ is the bond length, $b_{\rm max}$ is the maximum extension, and $k$
is the spring constant. Salt is added into the system. The salt molecules
dissociate into tetravalent cations (also called ``counterions'') and
monovalent anions (``coions'') in the solution. All the particles --- including
the monomers, counterions, and coions --- are modeled explicitly as spheres
that are described by the purely repulsive Lennard-Jones (LJ) potential
\begin{equation}
U_{LJ}(r)=\left\{\begin{array}{ll} 4\varepsilon_{LJ} \left[
\frac{\sigma}{r}\right)^{12}- \left(\frac{\sigma}{r}\right)^6 + \frac{1}{4} ] &
\mbox{, for } r \leq \root 6 \of 2 \sigma \\ 0 & \mbox{, for } r
> \root 6 \of 2 \sigma \end{array}\right.
\end{equation}
where $r$ is the separation distance; $\sigma$ and $\varepsilon_{\rm LJ}$
represent the diameter and the hardness of the LJ sphere, respectively. Since
the interaction between monomers is purely repulsive, our system corresponds to
a good solvent. Charged particles also interact with each other via the Coulomb
interaction
\begin{equation}
U_{\rm C}(r)= k_B T \lambda_B \frac{Z_i Z_j}{r}
\end{equation}
where $T$ is the temperature, $k_{B}$ is the Boltzmann constant, and $Z_i$ is
the charge valency of the $i$th particle. $\lambda_B=e^2/(4\pi \epsilon_0
\epsilon_r k_B T)$ is the Bjerrum length where $\epsilon_0$ is the vacuum
permittivity and $\epsilon_r$ is the relative dielectric constant of the
solvent. At the separation distance $r=\lambda_B$, the electrostatic energy
between two unit charges is exactly the thermal energy $k_{B} T$. The solvent
molecules are not modeled explicitly in the study. However, their effects are
incorporated implicitly through the following three ways: (1) the dielectric
constant $\epsilon_r$, which takes into account of the dielectric screening of
charge in the solvent medium, (2) the friction force $-m_i\zeta_i\vec{v}_i$,
which models the drag acting on particle $i$, proportional to the moving
velocity,  (3) the stochastic force $\vec{\eta}_i(t)$, which simulates the
thermal collisions of solvent molecules on the particle $i$. The equation of
motion is the Langevin equation,
\begin{equation}
m_i \ddot{\vec{r}}_i = -m_i\zeta_i \dot{\vec{r}}_i +\vec{F}_c +Z_i e E {\hat x}
+\vec{{\eta}}_i
\end{equation}
where $m_i$ is the particle mass, $\zeta_i$ is the friction coefficient, and
$\vec{F}_c=-\partial\,U/\partial\,\vec{r}_i$ is the conservative force. In
Langevin dynamics, the temperature is determined by the fluctuation-dissipation
theorem: $\left< \vec{\eta}_i(t) \cdot \vec{\eta}_j(t') \right> = 6 k_B T
m_i\zeta_i\delta_{ij} \delta(t-t')$. The external electric field is uniform and
exerts an electric force $Z_i e E {\hat x}$ on the particle $i$ in the
$x$-direction The system is placed in a periodic rectangular box.
Particle-particle particle-mesh Ewald sum is used to calculate the Coulomb
interaction~\cite{hockney88}.

We assume that all the particles have the identical mass $m$ and LJ parameters
$\sigma$ and $\varepsilon_{\rm LJ}$. We set $\varepsilon_{LJ}= 0.8333k_BT$,
$k=5.8333k_BT/ \sigma^2$, $b_{\rm max}=2 \sigma$, $\lambda_B=3 \sigma$, and
$\zeta_i=1 \tau^{-1}$, where $\tau=\sigma \sqrt{m/(k_BT)}$ is the time unit.
The chain length $N$ is varied and the monomer concentration $C_m$ is fixed at
$0.0003 \sigma^{-3}$, which describes a dilute polymer
solution~\cite{rubinstein03}. The simulation box is adjusted linearly with $N$
in $x$-direction, and has a volume of $1.3N \times 50.64 \times 50.64$. This
prevents chains from self-overlapping through the periodic boundary condition.
The salt concentration $C_s$ is varied from $C_s=0$ to $C_s=0.0006
\sigma^{-3}$, a range for the chain to exhibit the behavior of reentrant
condensation~\cite{hsiao06a, hsiao06b, hsiao06c}. The electric field strength
$E$ is varied over a wide range, from $E=0$ to $2k_BT/(e \sigma)$, to study its
effect on the properties of the PE system. The Langevin equation is integrated
by the Verlet algorithm. The integration time step $\Delta t$ is equal to
$0.005\tau$~\cite{note-lammps}. A pre-run phase takes about $10^7$ time steps
to bring the system into a stationary state, followed by a production-run phase
of $10^8$ time steps to cumulate data for analysis.  Since hydrodynamic
interaction is largely screened out under a typical electrophoretic
condition~\cite{long96,viovy00,tanaka02,tanaka03}, we neglect the hydrodynamic
interaction in this study. Recent simulations have demonstrated the validity of
this approximation when the chain length is not short ($N>20$)~\cite{grass09}.
We study the electrophoretic properties of chains at a fixed chain length
$N=48$ for the first 5 subsections of Sec.~III whereas in Sec.~III.F, $N$ is
varied from 12 to 768.  To simplify the notation, the value of a physical
quantity will be reported in the ($\sigma$, $m$, $k_B T$, $e$)-unit system in
the rest of the text. For example, the strength of electric field is described
in units of $k_BT/(e \sigma)$, the dipole moment is in units of $e\sigma$, and
the electrophoretic mobility is in units of $e\sigma^2/(\tau k_B T)$, and so
on.

\section{Results and Discussions} 
\subsection{Chain conformation in electric fields}
\label{Sec:conformation}
\begin{figure}[htbp]
\begin{center} \includegraphics[width=0.4\textwidth,angle=270]{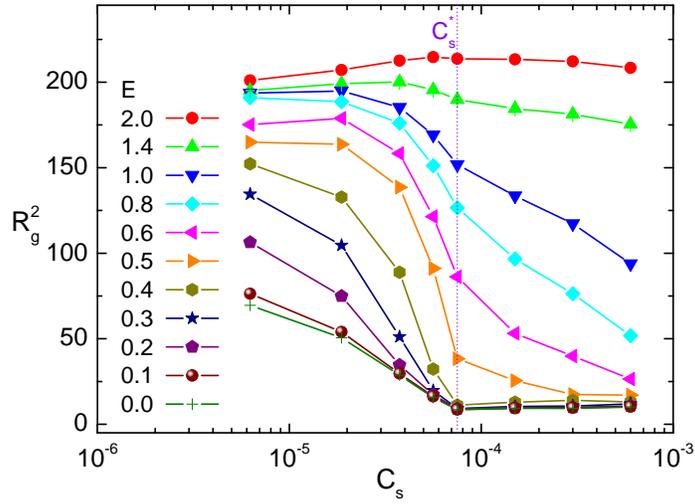}
\caption{Mean square radius of gyration $R_g^2$ as a function of tetravalent
salt concentration $C_s$ at different strength $E$ of electric field. The value
of $E$ can be read in the figure. The error bar of data in this paper is
smaller than the size of data symbol if it is not presented.}
\label{fig:RgvCsEX} \end{center}
\end{figure}

In this section, we study the conformation of PEs at different salt
concentrations $C_s$ under the action of electric fields. The chain length $N$
is 48.  We first calculated the mean square radius of gyration $R_g^2$, which
is used to characterize the chain size. The definition is given by $ R_g^2=
\sum_{i=1}^N \left<(\vec{r}_i-\vec{r}_{cm})^2\right>/N$ where $\vec{r}_{i}$ is
the position vector of the monomer $i$ and $\vec{r}_{cm}$ is the center of mass
of the chain. The results are plotted in Fig.~\ref{fig:RgvCsEX}. Each curve
shows how $R_g^2$ varies with $C_s$ at a given field strength $E$.

We see that in zero electric field, $R_g^2$ decreases with $C_s$ up to the salt
concentration $C_s^*=7.5\times 10^{-5}$. The value of $C_s^*$ is one fourth of
$C_m$, at which the amount of tetravalent cations in the solution are in charge
equivalence with the monomers on the chain.  The decrease of $R_g^2$ shows that
the chain collapses, which is due to ionic screening and to the bridging effect
induced by the added tetravalent salt. At $C_s=C_s^*$, the chain is nearly
neutralized by the condensation of tetravalent counterions.  In the region
$C_s>C_s^*$, $R_g^2$ becomes an increasing function of $C_s$. The chain size
reexpands slightly. The collapse and the reexpansion of the chains can be
regarded as a single-chain version of the PE precipitation and redissolution.
These phenomena are collectively called ``reentrant
condensation''~\cite{nguyen00}. Reentrant condensation for PEs has been
investigated in detail in theories~\cite{nguyen00, grosberg02, solis00,
solis01, solis02}, experiments~\cite{delacruz95, raspaud98}, and
simulations~\cite{hsiao06a, hsiao06b, hsiao06c, wei07}. 

Now let us focus on the cases in which the electric field is applied. All these
curves lie above the zero-field limit. This result demonstrates that the chain
is elongated by the applied electric field and therefore, takes a larger value
of $R_g^2$. The stronger the electric field, the larger the deviation from the
zero-field limit. However, this deviation also depends on $C_s$.  While
increased with $E$ at a given $C_s$ in the region $C_s<C_s^*$, $R_g^2$ is
basically unchanged with $E$ in the region $C_s>C_s^*$ for $E<0.4$.  When
$E>0.5$, the PEs no longer exhibits a compact structure at $C_s^*$.  The chain
size continues to shrink with $C_s$ for $C_s>C_s^*$.  This shrinkage suggests
that the excess of the tetravalent counterions in the solution helps the chains
to collapse, against the action of the electric fields.  In a very strong field
such as $E=2.0$, $R_g^2$ becomes basically a constant with $C_s$.  The chain
collapse by tetravalent salts is completely suppressed by the strong electric
field.
 
\begin{figure}[htbp]
\begin{center}
\includegraphics[width=0.4\textwidth,angle=270]{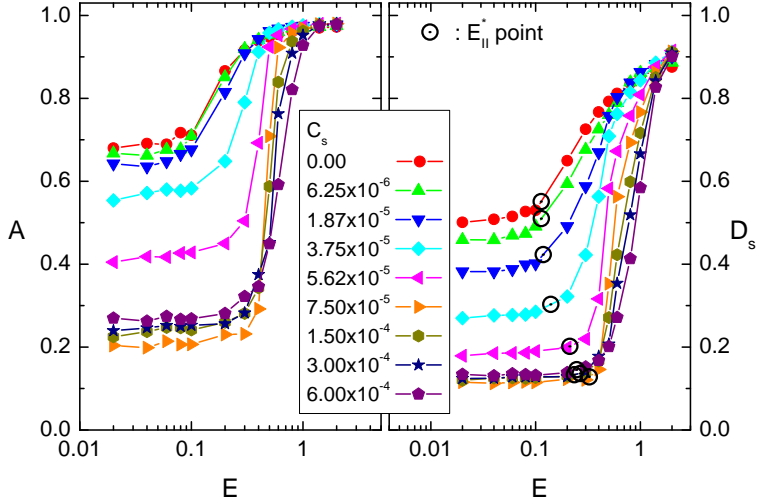}
\caption{\textcolor{\BLUE}{Asphericity $A$ and degree of chain unfolding $D_s$
vs. electric field strength $E$ at different $C_s$. The value of $C_s$ is
indicated in the legend.  The symbol ``$\odot$'' in the right panel denotes
the threshold electrical field (denoted by $E_{II}^*$ in the next section) at a
given $C_s$.}} \label{fig:AvESX} \end{center}
\end{figure}

We next calculated the asphericity $A$ of the chain in electric fields.  The
quantity  describes the degree of geometrical deformation away from a sphere
and can be used to characterize the chain conformation.  It is defined by
$A=\frac{1}{2}\left<\left( (\lambda_1-\lambda_2)^2 +(\lambda_2-\lambda_3)^2
+(\lambda_3-\lambda_1)^2 \right) /\left(\lambda_1+\lambda_2+\lambda_3
\right)^2\right>$ where $\lambda_1$, $\lambda_2$, and $\lambda_3$ are the three
eigenvalues of the gyration tensor of the chain. The tensor was calculated by
${\cal{T}}_{\alpha\beta}=\sum_{i=1}^{N} (\vec{r}_i-\vec{r}_{cm})_{\alpha}
(\vec{r}_i-\vec{r}_{cm})_{\beta}/N$, where the subscripts $\alpha$ and $\beta$
denote one of the three Cartesian components $x$, $y$, $z$ of the subscribed
vector, respectively. The value of $A$ ranges from 0 to 1. It is 0 for a
perfect sphere and 1 for a rod. For a random coiled chain, $A$ takes the value
$0.431$ obtained by simulations~\cite{bishop88}. The left panel of
Fig.~\ref{fig:AvESX} presents the asphericity $A$ of our chain as a function of
the electric field $E$.  Each curve denotes one case running at a given $C_s$.

The curves are horizontal lines when $E$ is small, showing that chain
distortion does not happen in the weak electric fields.  At $C_s=0.0$, the
value of $A$ is 0.68.  It is larger than 0.431, showing that the chain expands
more than a coil does.  We found that $A$ decreases with $C_s$ in this weak
$E$-region, and reaches a minimal value 0.20 at $C_s=C_s^*$.  The small value
of $A$ shows the formation of a compact globule structure.  When $C_s>C_s^*$,
$A$ begins to increase with $C_s$.  The result demonstrates a chain
reexpansion.  In the intermediate $E$ region between 0.1 and 1.0, $A$ exhibits
a drastic increase, showing the deformation of the chain by the electric field.
At large field $E=2.0$, the value of $A$ is around 1. Therefore, the chain is
deformed from its ``natural'' conformation in the zero field to a rodlike 
structure. 

\textcolor{\BLUE}{In the right panel of Fig.~\ref{fig:AvESX}, we show the
degree of chain unfolding $D_s$ as a function of $E$ at different $C_s$.  Here
$D_s$ is defined to be the ratio of the chain end-to-end distance $R_e$ to the
chain contour length $L_c$. Similar to $A$, $D_s$ displays a sigmoidal increase
when electric field is above some threshold value.  In very high electric
fields, the value of $D_s$ can be as large as 0.9, showing that the chain is
quasi-fully stretched.  We estimate the threshold field by the onset point, at
which $D_s$ increases $10\%$ from its zero-field limit.  The dependence of the
threshold field on $C_s$ can be clearly seen in the figure. }

\begin{figure}[htbp]
\begin{center}
\subfigure[]{\includegraphics[width=0.3\textwidth,angle=270]{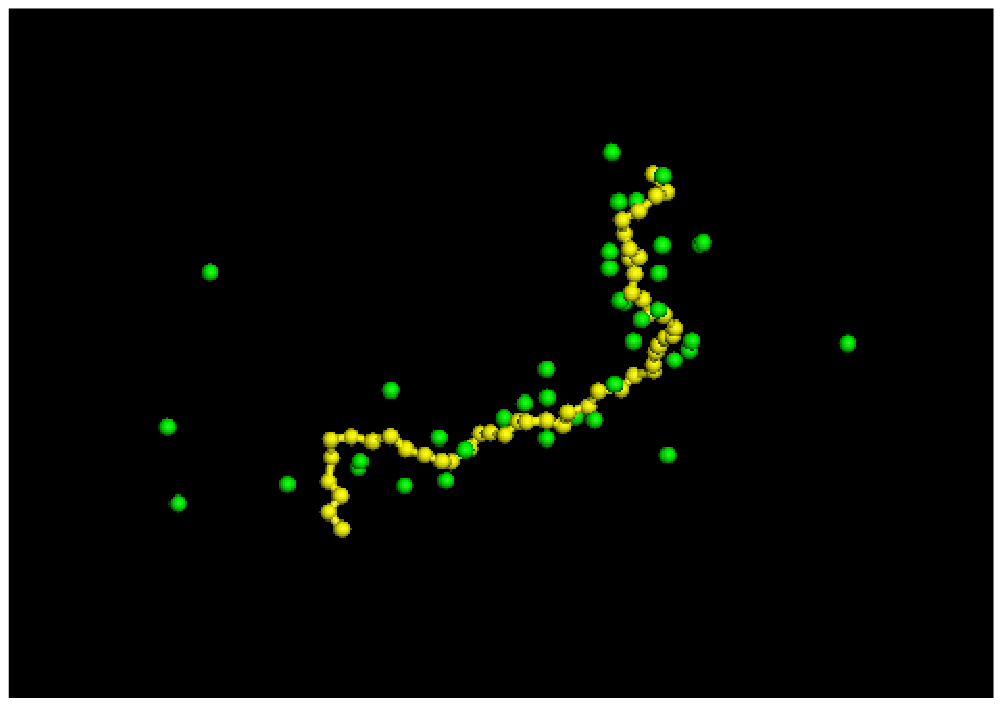}
\includegraphics[width=0.3\textwidth,angle=270]{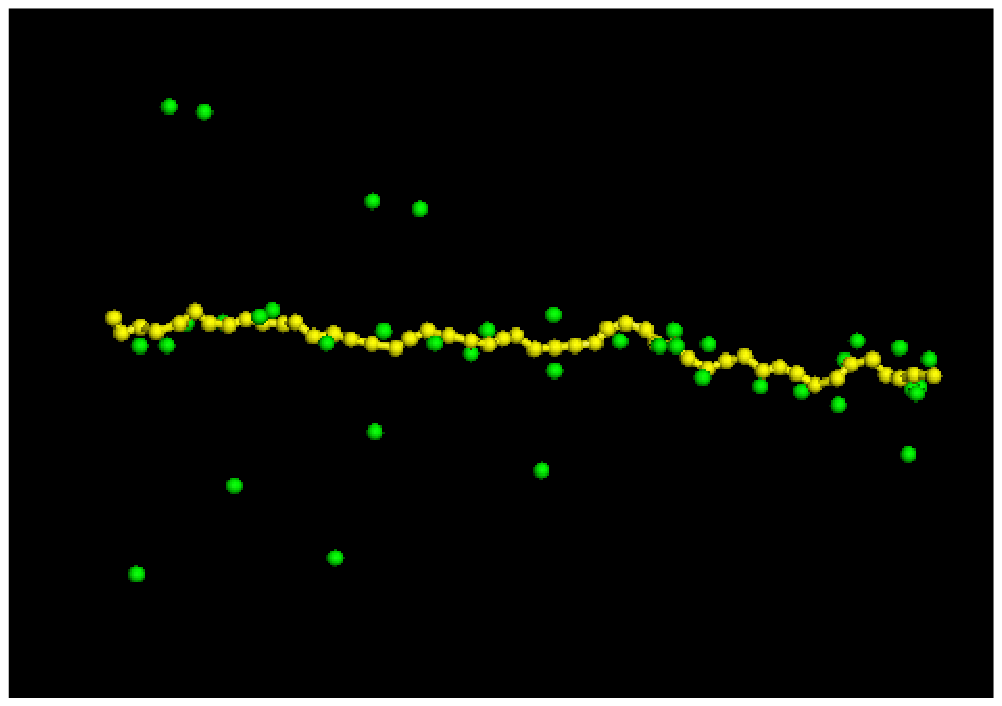}} \\
\subfigure[]{\includegraphics[width=0.3\textwidth,angle=270]{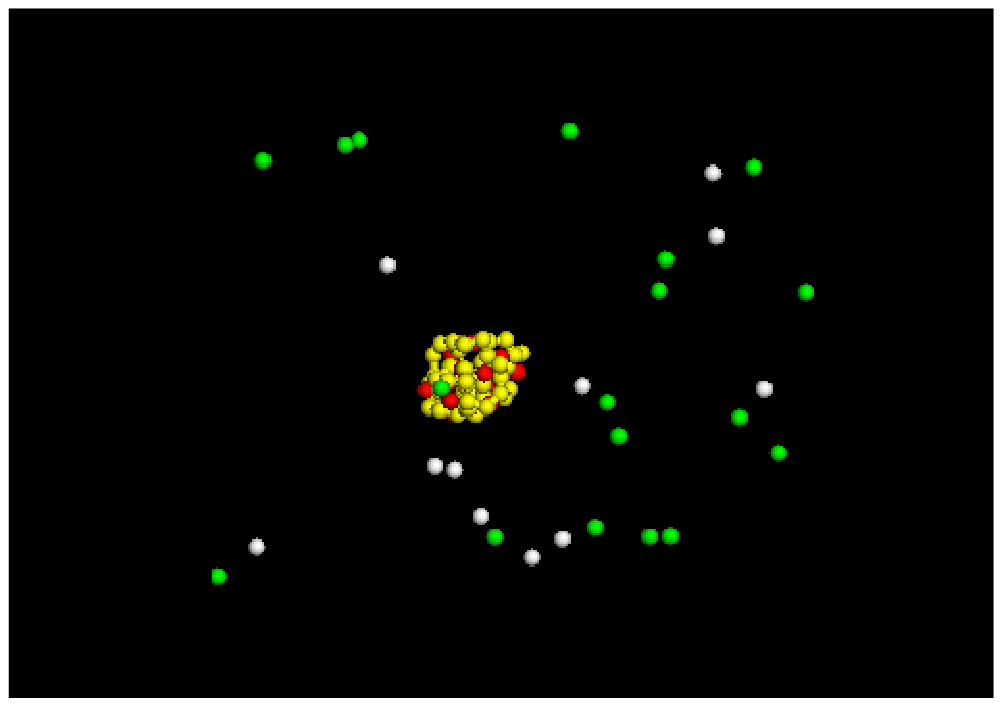}
\includegraphics[width=0.3\textwidth,angle=270]{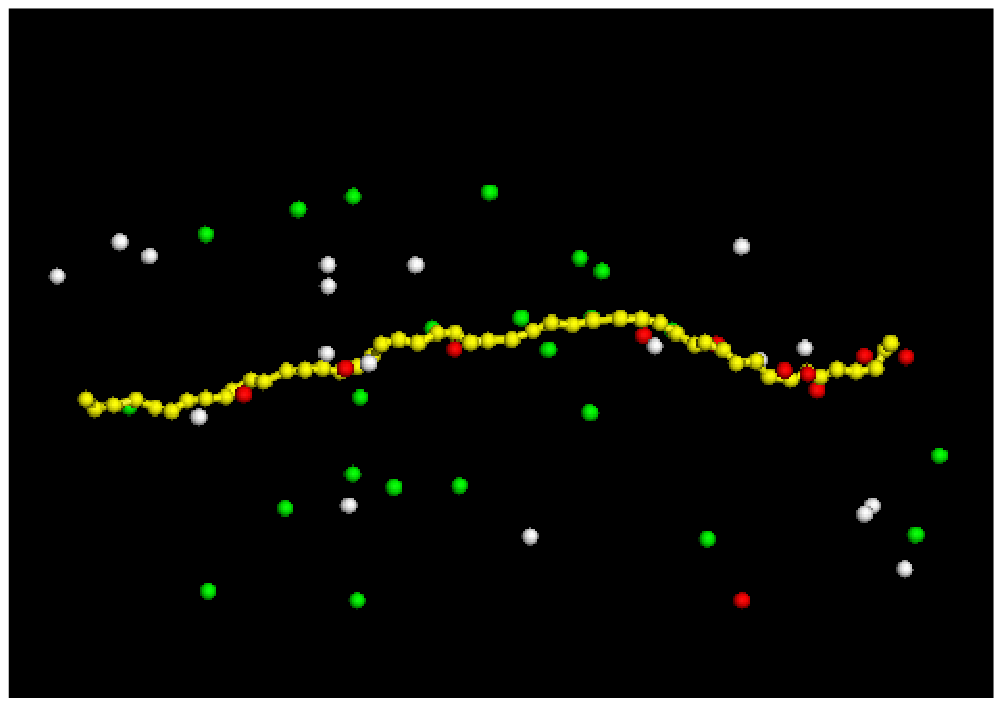}} \\
\subfigure[]{\includegraphics[width=0.3\textwidth,angle=270]{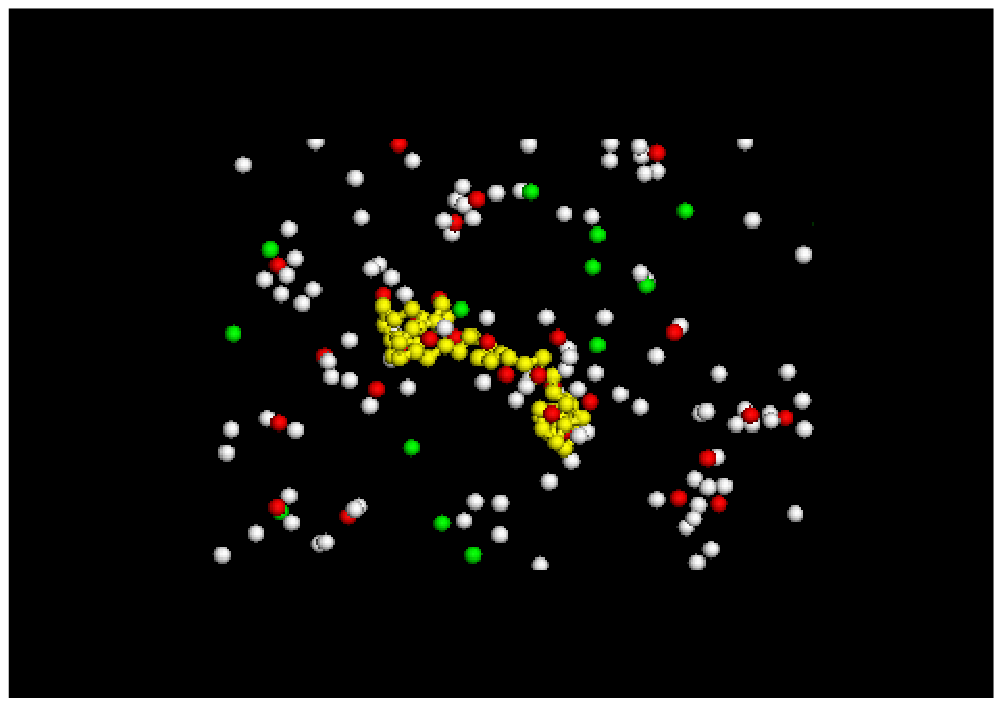}
\includegraphics[width=0.3\textwidth,angle=270]{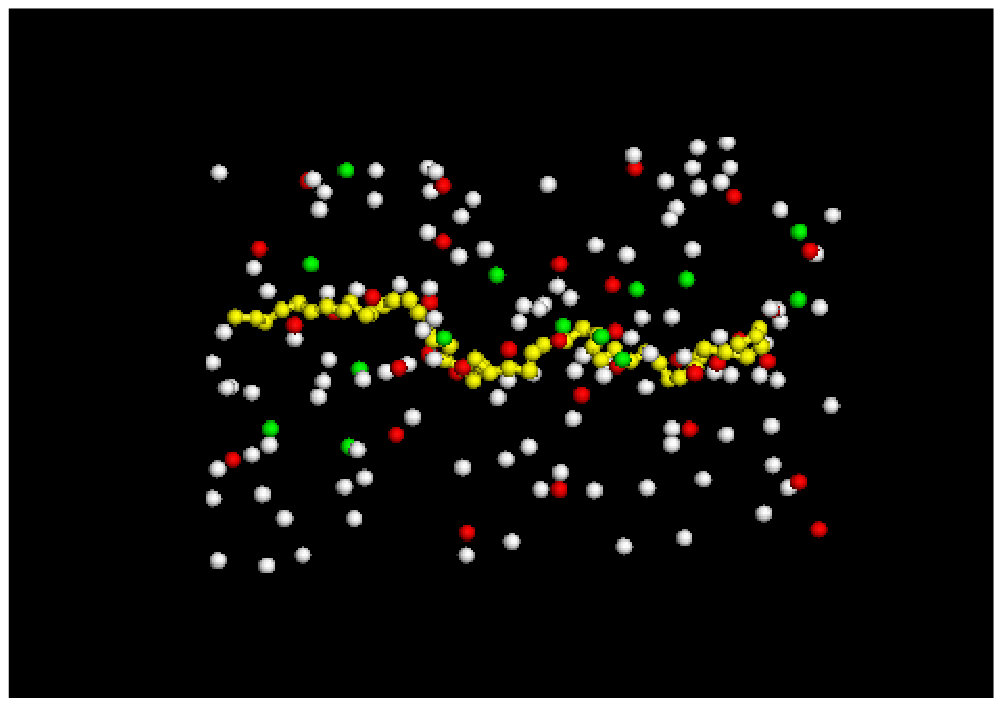}} \caption{(Color
online) Snapshots of simulation at (a) $C_s=0.0$, (b) $C_s=7.5\times10^{-5}$,
and (c) $C_s=6.0\times10^{-4}$.  In each panel, two pictures are presented: the
left one is the case in a weak field $E=0.02$ and the right one is in a strong
field $E=1.0$.  The PE is presented by a yellow bead-spring chain.  The
monovalent counterions, the tetravalent counterions, and the monovalent coions
are presented by green-colored, red-colored, and white-colored spheres. The
direction of electric field points toward the right direction.}
\label{fig:snapshots_dc} \end{center}
\end{figure}

To support the results of our calculation, we present in
Fig.~\ref{fig:snapshots_dc}, Panel (a), (b), and (c), the snapshots of
simulation at three salt concentrations $C_s=0$, $7.5 \times 10^{-5}$, and
$6.0\times 10^{-4}$, respectively. The three $C_s$ present the three cases with
the amount of the adding salt smaller than, equal to, and larger than the
equivalence point $C_s^*$.

The left picture in each panel of the figure shows the case subjected to a weak
electric field $E=0.02$, which is below the critical field $E^*$.  We can see
that the chain exhibits an expanded-coil structure in the salt-free solution
(Panel (a)), a compact globule structure at the equivalence point (Panel (b)),
and a less compact structure at the high salt concentration (Panel (c)),
consistent with the calculations of the asphericity in Fig.~\ref{fig:AvESX}.
The right picture in each panel presents the case in the strong electric field
$E=1.0$, which is above $E^*$. We can see that the chain unfolds to an
elongated structure. The chain size in Panel (a) is longer than in Panel (b),
and than in Panel (c), which agrees with the results in Fig.~\ref{fig:RgvCsEX}
where $R_g^2$ is a decreasing function over $C_s$ for $E=1.0$.  Moreover, we
can see that the elongated chain is aligned parallel to the field direction and
the condensed monovalent counterions (in Panel (a)) and the condensed
tetravalent counterions (in Panels (b) and (c)) cumulate more densely near the
chain end which follows the field direction (the right-hand end in
Fig.~\ref{fig:snapshots_dc} figure). It shows that the chain is polarized in
the electric field and possesses a dipole moment pointing to the field
direction to reduce the system energy. 

This dipole moment is precisely what causes the chain to unfold.  The electric
field causes an inhomogeneous distribution of counterions to condense on the
chain. Due to this inhomogeneity, the electric force acting on the two sides of
the chain is not balanced, which gives an effective tension along the chain.
The chain is unfolded if this tension is strong enough to break the
electrostatic binding between monomers by the condensed counterions.

\subsection{Polarization and critical electric field}
\label{sec:determineE}
The previous section discussed the existence of a critical field $E^*$, beyond
which a PE is aligned along the field direction and drastically unfolded. To
understand the unfolding mechanism in details, consider the polarization of a
chain of polymerization $N=48$ in electric fields.

\begin{figure}[htbp]
\begin{center}
\includegraphics[width=0.4\textwidth,angle=270]{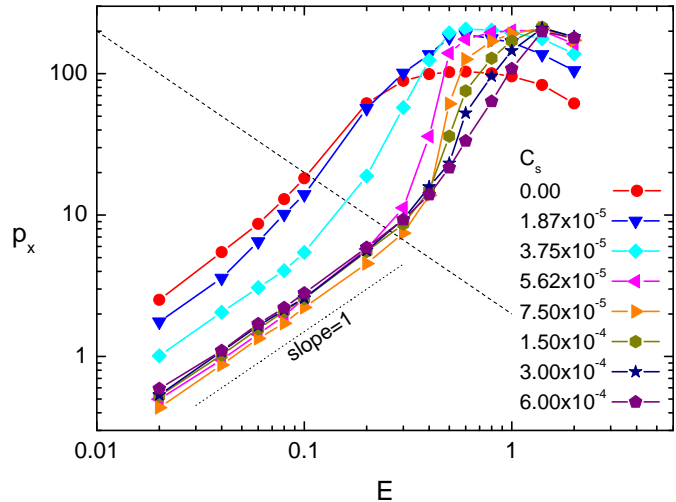}
\caption{Dipole moment $p_x$ as a function of $E$. Each curve corresponds to a
salt concentration whose value can be read in the figure. The dashed line is
the equation $p_x=2k_BT/E^*$ and the dotted line is a reference line with the
slope equal to 1. } \label{fig:PxvE} \end{center}
\end{figure}

The energy of the electric field stored in a dielectric material is
$W=(1/2)\int \vec{D}\cdot\vec{E} dV$ where $\vec{D}$ is the electric
displacement field. Since $\vec{D} = \epsilon_0\vec{E} +\vec{P}$ where
$\vec{P}$ is the polarization density, the energy $W$ can be written as a sum
of two terms: the first is $W_0=(1/2)\int \epsilon_0 |\vec{E}|^2 dV$, which
represents the field energy in vacuum, and the second is $W_{pol}=(1/2)\int
\vec{P}\cdot\vec{E}$, which denotes the polarization energy. In our study, the
polarization energy is calculated by $W_{pol}=\vec{p}\cdot\vec{E}/2$ where
$\vec{p}$ is the induced dipole moment of the PE-ion complex. For an unfolding
event to occur, the polarization energy $W_{pol}$ is larger than the thermal
fluctuation energy $k_BT$, according to Netz~\cite{netz03a, netz03b}.  The
dipole moment of the PE complex can be calculated by
\begin{equation}
\vec{p}=\sum_{i \in {\rm PE-complex}} Z_i e(\vec{r}_i-\vec{r}_{cm})
\end{equation}
where $\vec{r}_i$ is the position vector of the $i$th particle and $i$ runs
over the entire PE complex (both the monomers and the condensed ions), and
$\vec{r}_{cm}$ is the center of mass of the entire complex. There exists no
definite way to define a PE complex. In this study, we primitively define the
PE complex by a constant threshold distance $r_t$: a PE complex comprises the
chain itself and the ions with the distance to the chain smaller than $r_t$.
These ions are said to condense on the chain. We chose $r_t=\lambda_B$. This
criterion has been used previously to study ion condensation on a rigid
chain~\cite{manning69} but other definitions can be used~\cite{grass09}.  Since
the polarization will occur in the field direction, we calculated here the
$x$-component of the dipole moment, $p_x$.  The results are presented in
Fig.~\ref{fig:PxvE} as a function of $E$ for different $C_s$.  The criterion
for chain unfolding is that $p_x \ge 2k_BT/E$, and a dashed line demarcates the
boundary of the two regions in the figure.  $W_{pol}$ is larger than $k_BT$
above the line, whereas smaller below the line.
   
The points of $p_x$ fall on a straight line in the log-log plot in the weak
field region $E<2k_BT/p_x$. The slope of the line is 1, which follows exactly
linear response theory, $p_x=\alpha E$.  The polarizability $\alpha$ is
directly related to the height of the line in the figure.  $\alpha$ takes a
large value for a non-collapsed chain, and is smallest for a collapsed chain at
$C_s=C_s^*$.  In the field region above the dashed line, $p_x$ deviates from
this simple power law. Approximately above the dashed line, the curves grow
faster than the linear dependence but then curve down, exhibiting a hook-like
behavior. 

\textcolor{\BLUE}{In the weak field region, the chain conformation is not
perturbed by the applied electric field. However, the displacement of the
condensed ions on the chain is still possible. It is hence the ion displacement
that is responsible for the formation of the dipole moment and results in the
linear response behavior.  When the electric field is strong enough to unfold
the chain, the chain elongation provides the further possibility for the
condensed ions to migrate on the chain~\cite{hsiao11}. Consequently, the dipole
moment acquires a value larger than predicted by linear response theory.  In
very strong electric fields, the chain reaches its maximum extension. Further
elongation of the chain becomes impossible. Increasing the electric field can
cause condensed ions on the chain to be stripped off. The number of the
condensed ions decreases. As a consequence, the dipole moment decreases,
resulting in the hook-like curve.}

\begin{figure}[htbp]
\begin{center} \includegraphics[width=0.4\textwidth,angle=270]{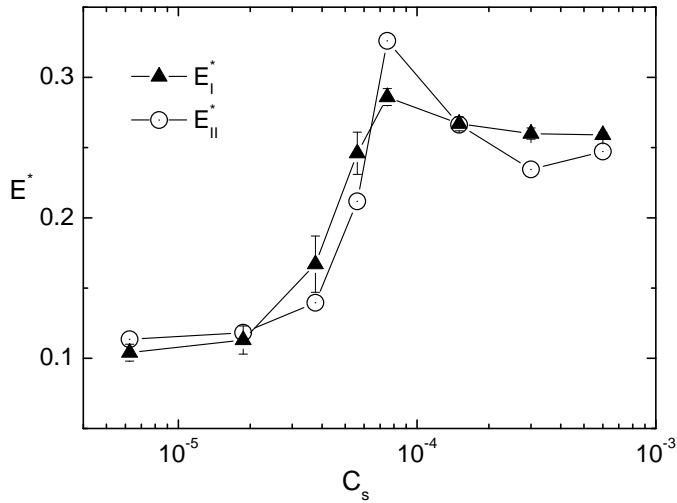}
\caption{Critical electric field $E^*$ as a function of $C_s$. $E^*$ is
determined by two different methods: $E^*_I$ from the polarization energy and
$E^*_{II}$ from the onset point of $D_s$. } \label{fig:cEvCs} \end{center}
\end{figure}

\begin{figure}[htbp]
\begin{center}
\includegraphics[width=0.4\textwidth,angle=270]{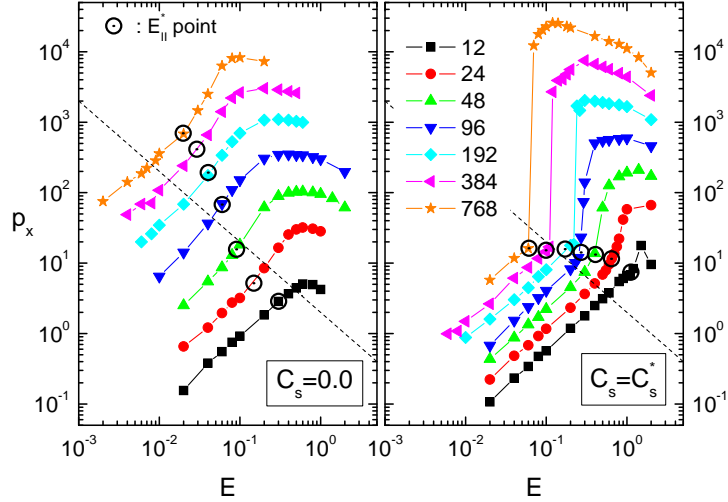} \caption{
\textcolor{\BLUE}{Dipole moment $p_x$ as a function of $E$ at $C_s=0$ and
$C_s=C_s^*$.  Chain length $N$ is indicated in the right panel.  The dashed line
defines the equation $p_x=2k_BT/E^*$.  The symbol ``$\odot$'' denotes 
the dipole moment at $E=E_{II}^*$.}} \label{fig:P1Nxx_PxE_rc3} \end{center}
\end{figure}

The critical electric field $E^*$ is estimated as the intersection of the
simulated $p_x$ curves with the unfolding transition $p_x=2k_BT/E^*$.  The
obtained critical field is reported in Fig.~\ref{fig:cEvCs} as a function of
$C_s$ and denoted $E^*_I$ (for reasons that will be clear momentarily).
We can see that $E^*_I$ increases with $C_s$ and reaches the maximum value at
$C_s=C_s^*$, and then, decreases slightly. This behavior is consistent with
what we have observed in Fig.~\ref{fig:RgvCsEX}. The chain is easier to stretch
in the low-salt region ($C_s<C_s^*$) than in the high-salt region
($C_s>C_s^*$). \textcolor{\BLUE}{To verify that the chain begins unfolding at
$E=E^*_I$, the threshold field obtained from the onset increasing of $D_s$ in
Fig.~\ref{fig:AvESX} is plotted and denoted by $E^*_{II}$. The consistency
between $E^*_{I}$ and $E^*_{II}$ data seems to suggest that chain unfolding
occurs as when linear response theory no longer holds. Nonetheless, further
verification through varying the chain length from $N=12$ to $768$ shows that
$E_I^*$ is not always situated at the break-down of the linear dependence,
as shown in Fig.~\ref{fig:P1Nxx_PxE_rc3}.}

\textcolor{\BLUE}{Indicating the onset-critical point on the curve,
we clearly see that the dipole moment deviates from the linear
behavior at $E=E_{II}^*$, but not at $E=E_{I}^*$.  Therefore, the criterion
$W_{pol}\ge k_BT$ for chain unfolding is too simple to be precise.  A real
unfolding point can happen with the polarization energy larger or smaller than
$k_BT$, depending on both the chain length and salt concentration.  It
necessitates a fundamental understanding of the unfolding mechanism in the
future to set up a correct criterion for the problem. The results shown here
demonstrate that chain unfolding is tightly connected with the polarization 
change. }

\subsection{Electrophoretic mobility}
We have seen that PE-ion complexes exhibit a drastic unfolding transition when
the electric field is stronger than $E^*$. One pertinent question is whether
the mobility of the chain shows a drastic change too, accompanying the
conformational transition. Also, it is very important to know the mobility of
ions, especially the condensed multivalent counterions, because these ions
play a decisive role in determination of the chain conformations. 

\begin{figure}[htbp]
\begin{center}
\subfigure[]{\includegraphics[width=0.35\textwidth,angle=270]{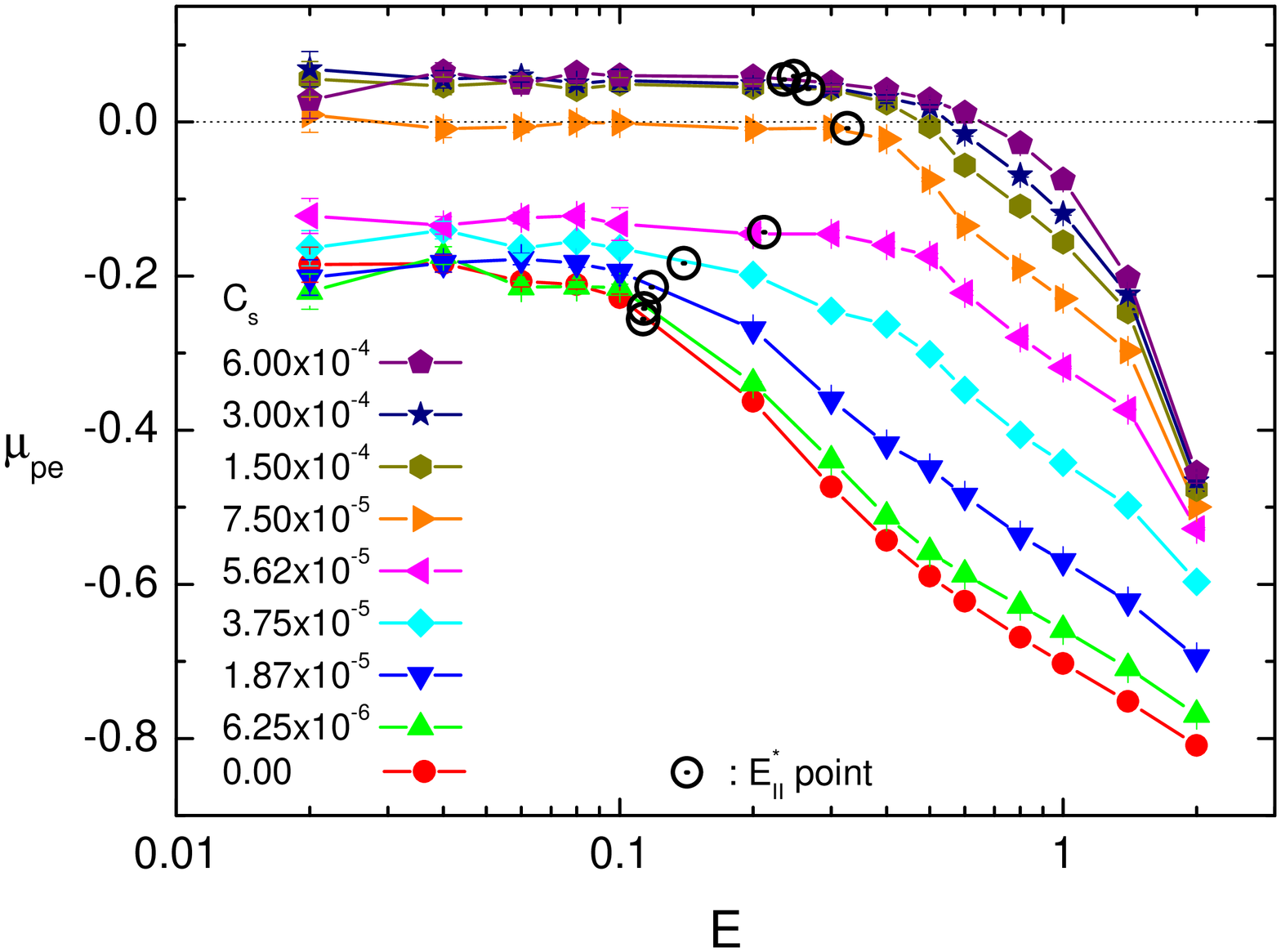}}
\subfigure[]{\includegraphics[width=0.35\textwidth,angle=270]{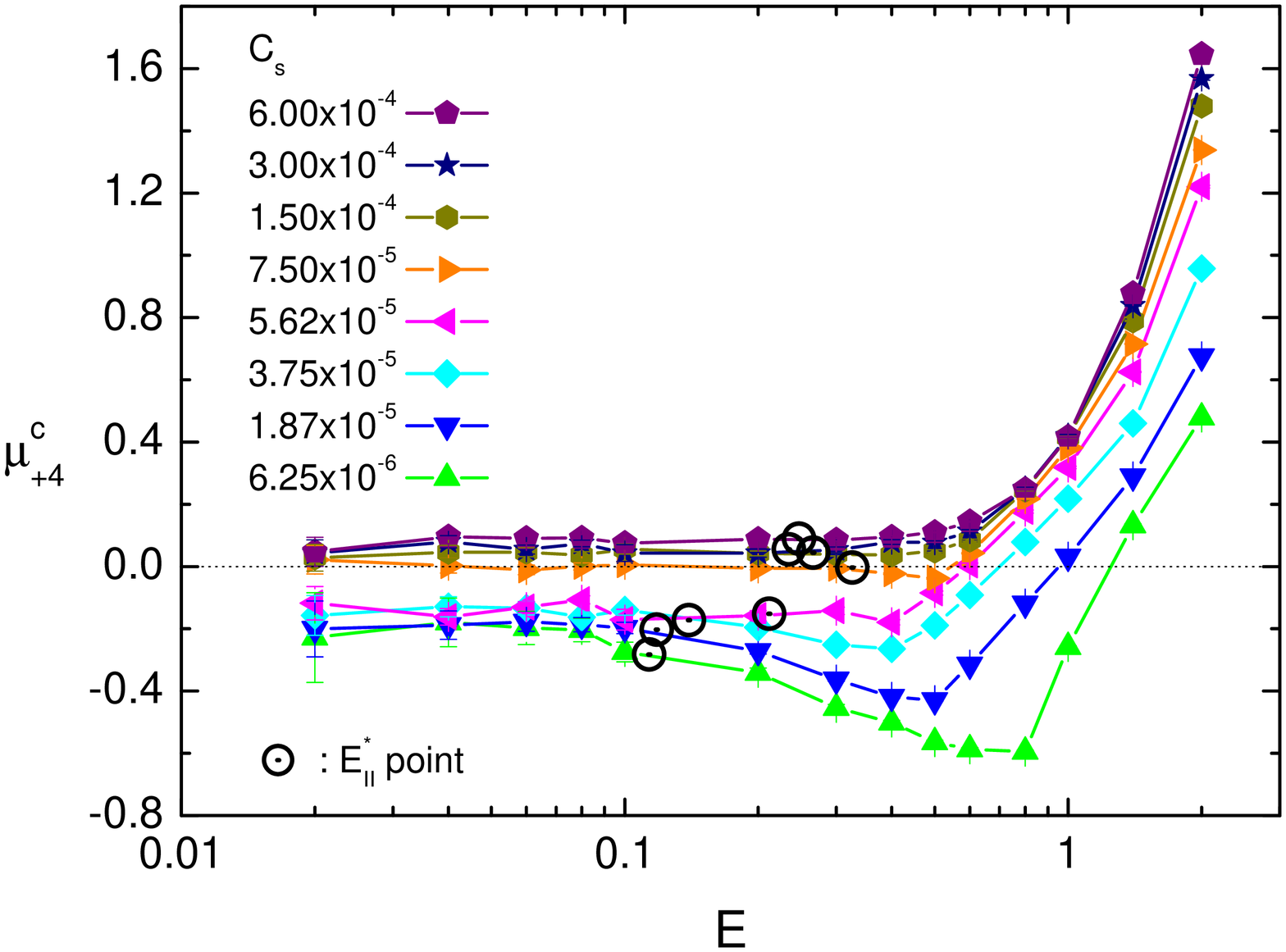}}
\subfigure[]{\includegraphics[width=0.35\textwidth,angle=270]{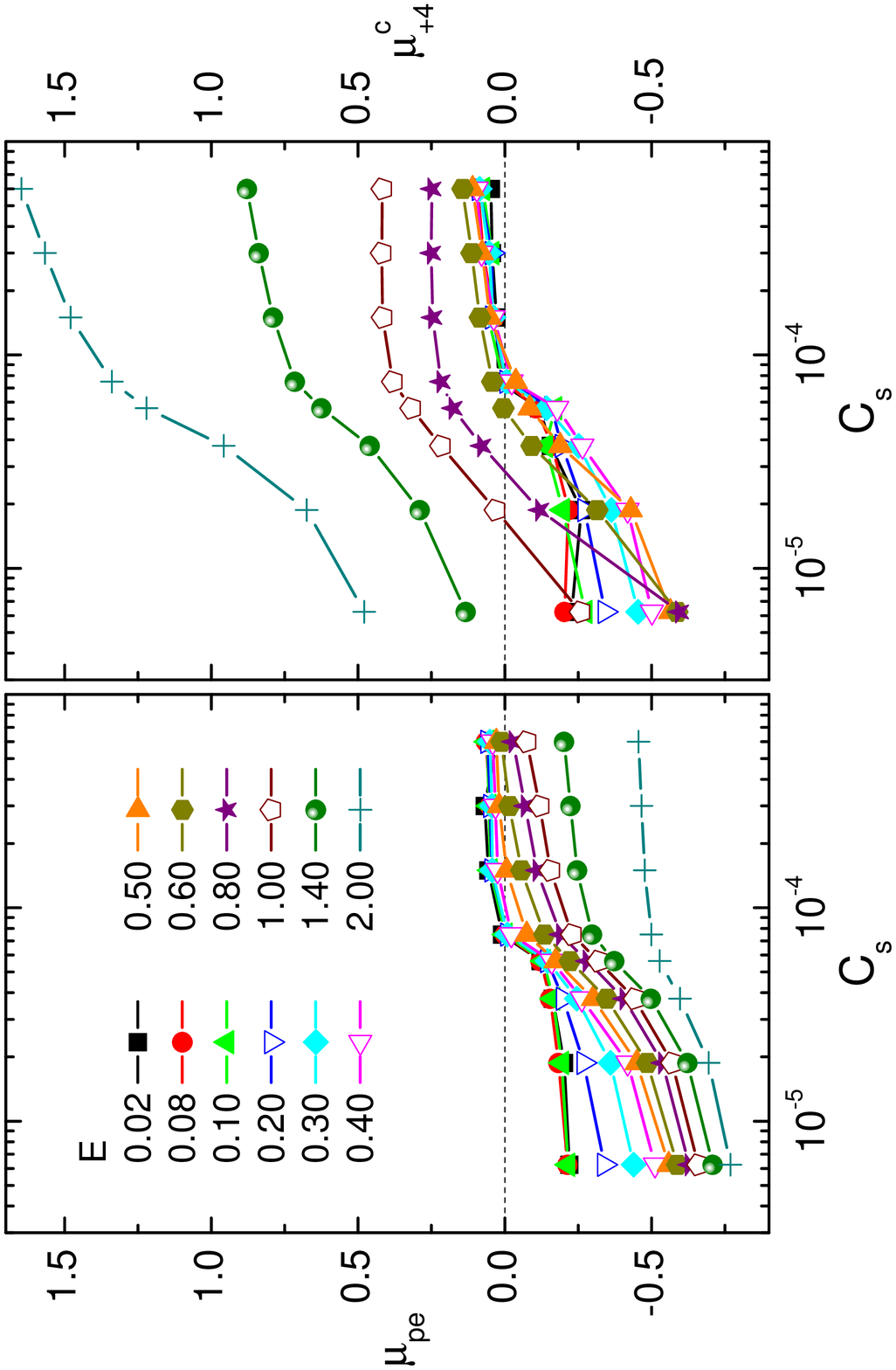}}
\caption{Mobility of (a) chain $\mu_{pe}$ and (b) the condensed tetravalent
counterions $\mu_{+4}^c$ as a function of $E$.  Each curve denotes one case
running at a specific $C_s$. The value of $C_s$ can be read in the figure.  The
critical field $E^*_{II}$ is indicated on the corresponding curve by the symbol
``$\odot$''. (c) $\mu_{pe}$ and $\mu_{+4}^c$ vs.~$C_s$ in different $E$ fields,
replotted from the data of (a) and (b). The value of $E$ is indicated in the
left panel.} \label{fig:mobility} \end{center}
\end{figure}

The mobility of the chain and of the condensed tetravalent counterions are
defined as $\mu_{pe}=v_{pe}/E$ and $\mu_{+4}^c=v_{+4}^c/E$, respectively, where
$v_{pe}$ and $v_{+4}^c$ are, in turn, the velocities of the chain and the
condensed tetravalent counterions in the field direction. The results are
presented in Fig.~\ref{fig:mobility}, Panels (a) and (b), for $\mu_{pe}$ and
$\mu_{+4}^c$, respectively, as a function of $E$ for different $C_s$.  The sign
of the obtained mobility can be positive or negative, denoting the moving
direction of the studied object toward the $+x$- or the $-x$-direction. To
investigate the relationship between the mobility change and the conformational
transition, we have plotted on the curves of the data the corresponding
critical field $E^*$ (the $E^*_{II}$) by the symbol ``$\odot$''.

We can see in Fig.~\ref{fig:mobility}(a) that when $E<E^*$, $\mu_{pe}$ is a
constant and depends on the salt concentration.  For $C_s<C_s^*$, the value of
$\mu_{pe}$ is smaller than zero because the effective chain charge is negative,
the same sign as the bare chain charge.  Increasing $C_s$ leads to a decrease
of $|\mu_{pe}|$. The absolute net chain charge decreases due to the
condensation of the counterions on the chain backbone.  At $C_s=C_s^*$,
$\mu_{pe}$ is approximately zero. This is because the total charge of the
tetravalent counterions in the solution is equivalent to the charge of the PE.
The chain is effectively neutralized upon the condensation of these ions.  If
$C_s$ is increased above $C_s^*$, $\mu_{pe}$ becomes positive. The chain now
moves in the $+x$-direction and the effective charge of the PE-ion complex is
positive. In other words, a charge inversion occurs. Charge inversion induced
by multivalent salt has been observed in experiments~\cite{kruyt49}.
Nonetheless, researchers continue to put their efforts on this study area for a
full understanding of the underlying mechanism~\cite{elimelech90, defrutos01,
hsiao08a}.  When the electric field is strong $E>E^*$, $\mu_{pe}$ is no longer
a constant but monotonically decreases with $E$. The fact that the effective
chain charge becomes more negative suggests that a strong electric field strips
the condensed counterions off the chain.  A detailed study concerning the
number of the ions condensing on the chain will be presented in the next
section. Our results reveal that charge inversion can be suppressed by strong
electric fields.  Comparing mobility with the conformational change
demonstrates that chain mobility and unfolding are closely related.

Fig.~\ref{fig:mobility}(b) shows the mean mobility of the tetravalent
counterions $\mu_{+4}^c$ condensed on the chain.  \textcolor{\RED}{ We observed
that $\mu_{+4}^c$ is approximately equal to $\mu_{pe}$ when $E$ is small.  We
recall that the mobility of tetravalent counterions is positive in a free
solution, but Fig.~\ref{fig:mobility}(b) shows that $\mu_{+4}^c <0$ when
$C_s<C_s^*$.  A negative $\mu_{+4}^c$ clearly demonstrates that the
electrostatic interaction between tetravalent counterions and monomers sets a
strong constraint on the ions. The ions are dragged along with the chain and
consequently, $\mu_{+4}^c\simeq \mu_{pe}$.  When $E$ is strong enough to break
the constraint, $\mu_{+4}^c$ deviates from $\mu_{pe}$.  In very strong electric
fields, these ions glide along the chain surface in the $+x$-direction;
$\mu_{+4}^c$ thus takes a positive value. The mobility difference between the
chains and the condensed ions suggests that ion condensation must take place in
a dynamic way. The condensed ions glide on the chain and are eventually
stripped off the chain end. The ions from the bulk solution continually
condense onto the chain from the other end. The process repeats continuously
and a balance is established.}

\textcolor{\RED}{Please notice that the field required to cause $\mu_{+4}^c$ to
deviate from $\mu_{pe}$ is stronger than $E^*$. This suggests that the chain
unfolding occurs before the ions are able to glide along the chain. Therefore,
there is a field region in which the chain has unfolded but the condensed
tetravalent counterions are too tightly bound to glide.  Since $\mu_{pe}$
becomes more and more negative, so does $\mu_{+4}^c$ before shooting up in
stronger $E$ fields.  Consequently, the $\mu_{+4}^c$ curve exhibits a minimum.
For $C_s > C_s^*$, the binding between condensed tetravalent counterions and
monomers becomes effectively weaker because of overcharging (see
Fig.~\ref{fig:condIons_E_Sxxx}(b)). The field region where the chain unfolds
without ion gliding shrinks as $C_s$ is increases.  The minimum disappears and
$\mu_{+4}^c$ grows monotonically with $E$.  Fig.~\ref{fig:mobility}(c) shows
the mobility presented as a function of salt concentration in different field
strength. As we can see, $\mu_{pe}$ and $\mu_{+4}^c$ both increase with $C_s$.
Moreover, $\mu_{pe}$ is roughly equal to $\mu_{+4}^c$ over the studied $C_s$
for $E\le0.4$. For $E>0.4$, very different behavior is observed.  The mobility
of the tetravalent ions increases much faster than $\mu_{pe}$, and eventually
becomes entirely positive at very strong $E$ while $\mu_{pe}$ remains
negative.}

\textcolor{\RED}{We remark that in this study, each data point of the mobility
is calculated under the action of an electric field at a given strength, once
the system reaches a steady state.  Figs.~\ref{fig:mobility}(a) and (b) are not
obtained by sweeping the DC field strength at a constant rate.  Therefore, no
hysteresis is displayed.  Our results report the mobilities in a stationary,
pseudo-equilibrium condition.  For simulations done by sweeping the field
strength, one would expect the occurrence of a hysteretic behavior in mobility,
and also in chain size, if the sweeping rate is fast. There exists two relevant
characteristic times in chain conformational transitions: the first is the
chain unfolding time to transition from a coiled or globular chain to an
elongated one, and the other is the collapse time to relax from an elongated
chain to a coil or globule.  The unfolding time is generally longer than the
collapse time because additional time is needed to disentangle a coiled or
globular chain.  The chain size (and therefore mobility) will follow a
different path when sweeping from a weak DC field to a strong one and then from
a strong field to a weak one, presumed that the total sweeping time would be
comparable to the two characteristic times. This deserves further investigation
in the future.  A counterpart has been shown in experiments studying chain-size
hysteresis when AC field frequency is changed at a fixed field
amplitude~\cite{wang10}. }

\subsection{Number of condensed ions and effective chain charge in electric
fields}
\begin{figure}[htbp]
\begin{center}
\subfigure[]{\includegraphics[width=0.35\textwidth,angle=270]{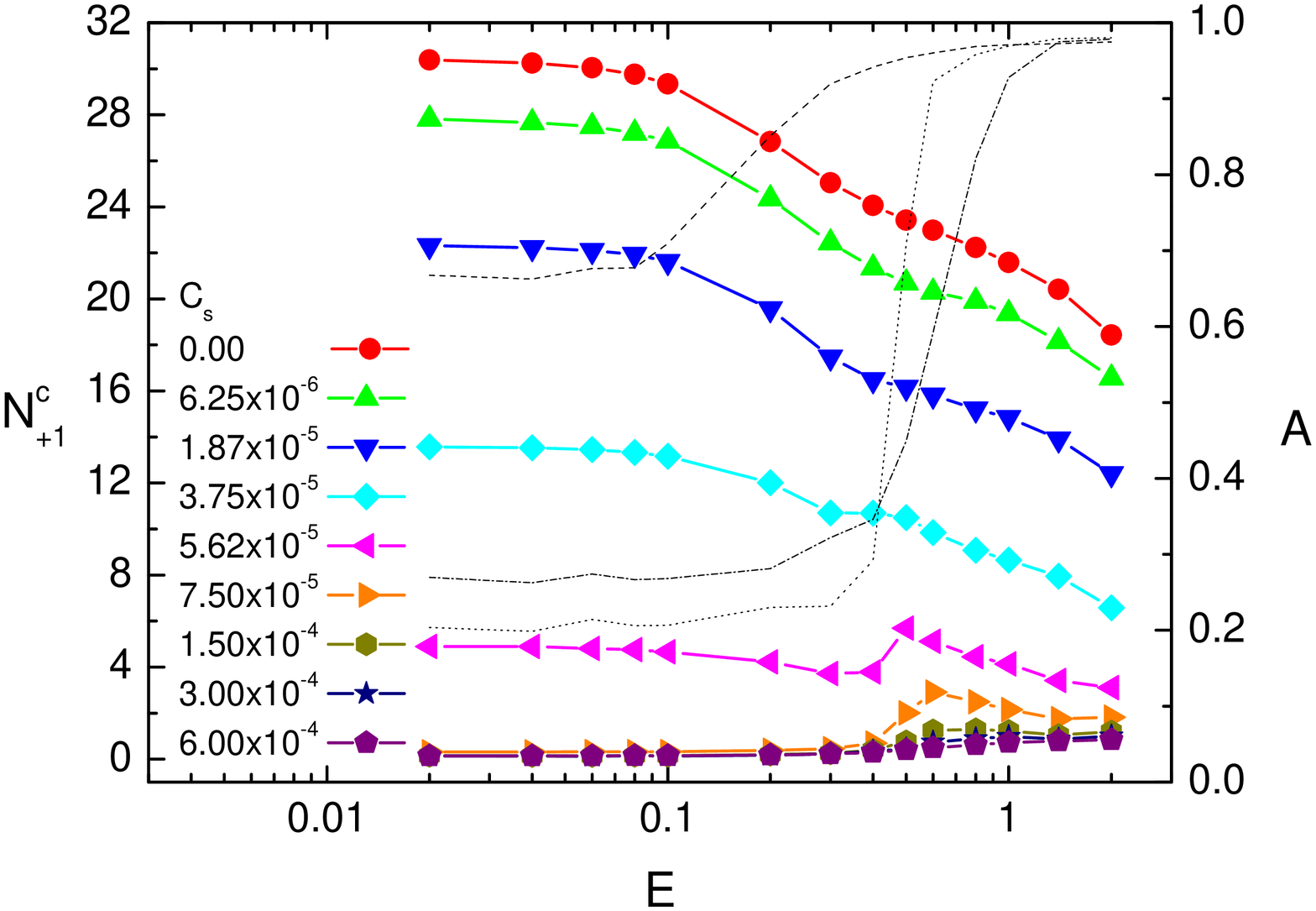}}
\subfigure[]{\includegraphics[width=0.35\textwidth,angle=270]{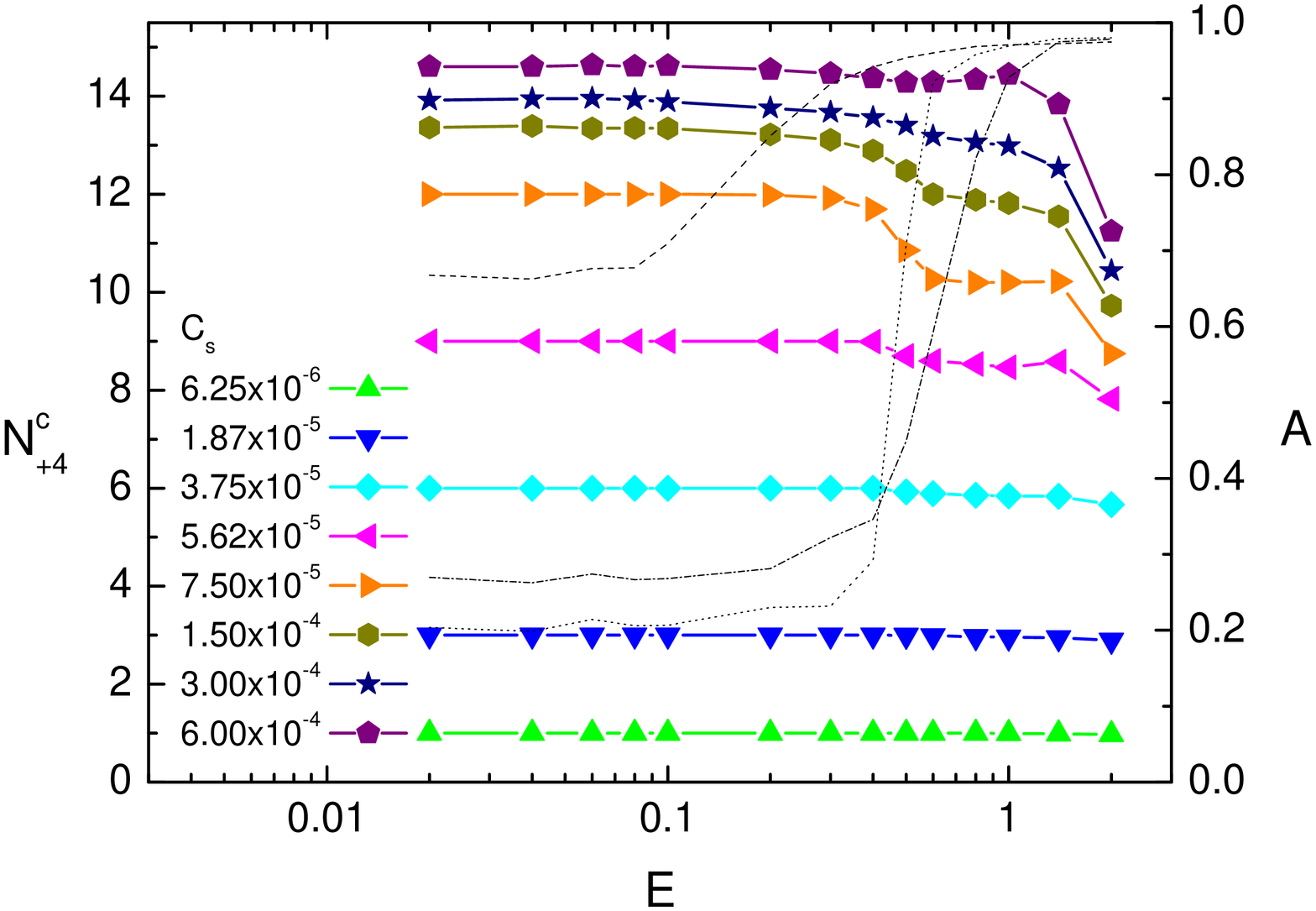}}
\subfigure[]{\includegraphics[width=0.35\textwidth,angle=270]{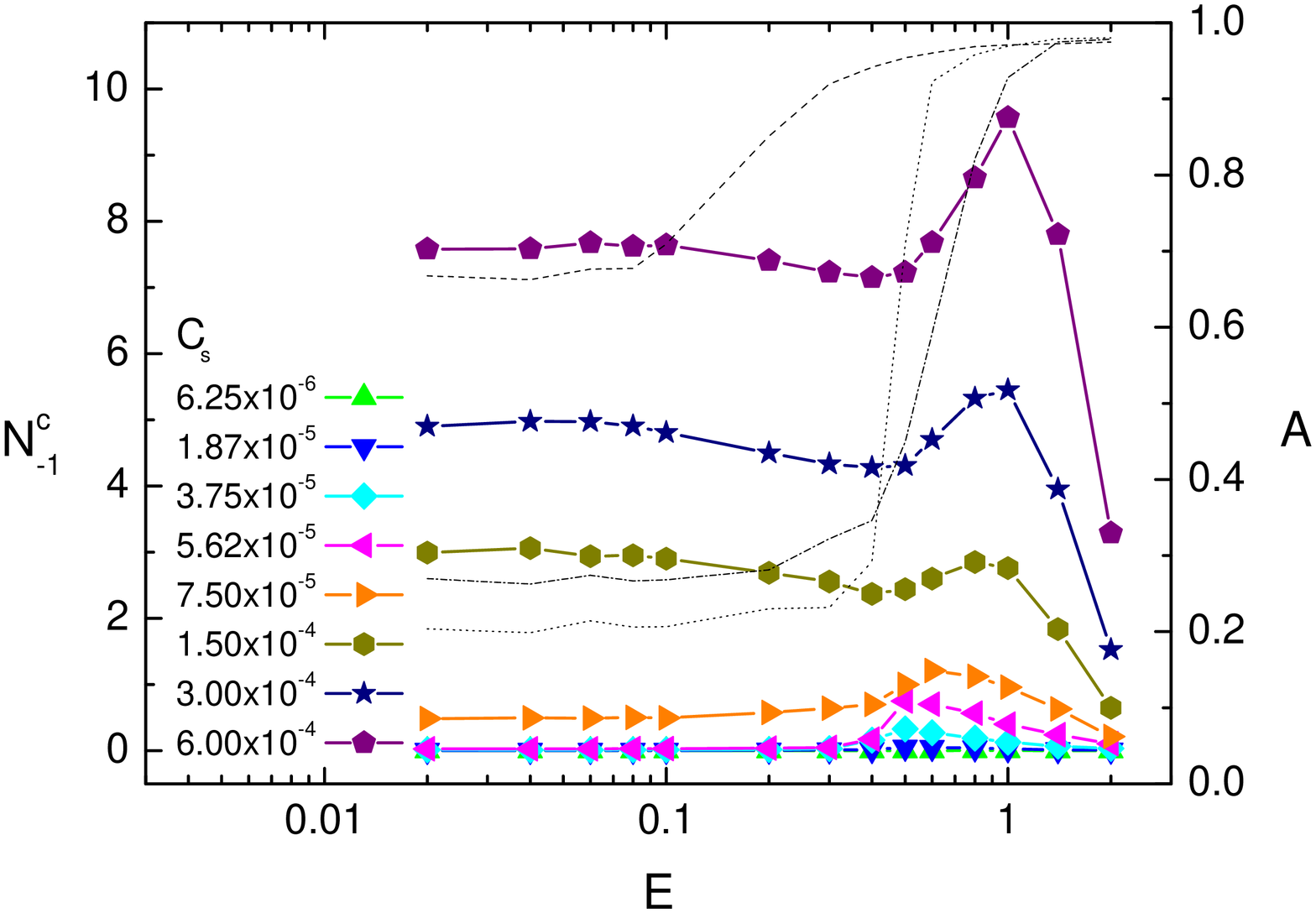}}
\caption{Number of condensed (a) monovalent counterions $N_{+1}^c$, (b)
tetravalent counterions $N_{+4}^c$, and (c) coions $N_{-1}^c$ as a function of
$E$.  The salt concentration can be read in the figures.  In each panel, the
chain asphericity $A$ at $C_s=6.25\times 10^{-6}$ (dashed curve), $7.5\times
10^{-5}$ (dotted curve), and $6.0\times 10^{-4}$ (dash-dotted curve) are
plotted to illustrate the chain conformational change.  The value of $A$ is
read from the right y-axis of the plot. } \label{fig:condIons_E_Sxxx}
\end{center}
\end{figure}

In this section, we study the number of ions condensed on the chain ($N=48$)
and effective chain charge in electric fields.  As in the previous section, an
ion is regarded to condense onto the chain if its distance to the chain is
smaller than the threshold distance $r_t=\lambda_B$.  Once the condensed ions
are identified, the effective chain charge can be calculated by summing the
charges of the condensed ions and the chain monomers.  Since there are three
types of ions in the simulation box, we treat each species independently.  The
results are presented in Fig.~\ref{fig:condIons_E_Sxxx} as a function of $E$
where $N_{+1}^c$, $N_{+4}^c$, and $N_{-1}^c$ are the numbers of the condensed
monovalent counterions, tetravalent counterions, and coions, respectively.  In
order to make comparison with the chain conformational transition, we have
plotted the chain asphericity $A$ at three representative $C_s$ in the figure.

We observe that $N_{+1}^c$ is  unperturbed by $E$, when $E<E^*$.  This
constant value decreases with $C_s$ and goes to zero when $C_s \ge C_s^*$:
There are no condensed monovalent counterions above the equivalence point.
When $E>E^*$, $N_{+1}^c$ shows a decreasing behavior with $E$ for $C_s\le 3.75
\times 10^{-5}$ but at $C_s=5.62\times 10^{-5}$ and $7.5\times 10^{-5}$, the
$N_{+1}^c$ exhibits a small hump.  For even higher $C_s$, the curve increases
slightly.  By comparing with the chain asphericity, we find that these
variations take place at the moment when the chain changes its conformation.
Therefore, the number of condensed ions is closely related to chain morphology.
How salt-induced chain conformation affects the ion condensation in zero
electric field has been studied in reference~\cite{wei09}. 

$N_{+1}^c$ at $C_s=0.0$ can be used to verify Manning condensation
theory~\cite{manning69}. The theory states that counterion condensation takes
place on a rigid PE while the mean distance per unit charge on the chain is
smaller than the Bjerrum length.  It results in an effective chain charge equal
to $-eL_c/\lambda_B$.  In our simulations, the mean distance before
condensation takes place is equal to the averaged bond length
$\left<b\right>=1.1$.  Hence Manning's theory predicts an effective chain
charge of $-17.3$. This is equivalent to 30.7 monovalent counterions condensing
on the chain.  Our simulations obtained $N_{+1}^c=30.4$, very close to the
prediction, although the chain is flexible and of finite length, which does not
follow the assumption of a rigid, infinitely long chain. This consistency shows
that the theory is a good approximation of flexible chains and also that the
choice for the condensation threshold $r_t=\lambda_B$ able to capture the
condensation phenomena.  

In Fig.~\ref{fig:condIons_E_Sxxx}(b), we observed that $N_{+4}^c$ remains
constant for a much wider range of electric field than $N_{+1}^c$.  For
example, in $C_s \le 3.75 \times 10^{-5}$, this number varies very little over
the studied field strengths, although the electrophoretic mobility (in
Fig.~\ref{fig:mobility}) does show some variation.  Therefore, the decrease of
$\mu_{pe}$ in this salt region is directly related to the stripping-off of
condensed monovalent counterions from the chain.  Moreover, we observed that
$N_{+4}^c$ exhibits a two-step decrease at $C_s \simeq C_s^*$.  By referring to
the corresponding $A$ curve, we can find that the first plateau appears in the
weak field region where the chain exhibits a unperturbed, globular structure
and the condensed tetravalent counterions are wrapped within the twisted
chain~\cite{hsiao06a, hsiao06b}.  The second, small plateau occurs when $E$ is
strong and the chain is completely stretched with $A\simeq 1$.  In this case,
the number of the condensed tetravalent counterions maintains a constant that
is smaller than the globular chain value.  If the applied field is very strong,
such as $E>1$, the tetravalent counterions can be stripped off the chain
violently; as a consequence, $N_{+4}^c$ decreases.  The two-step plateaus are
smeared out when $C_s$ is high.  It is because a large number of the
tetravalent counterions appears in the solution, obscuring the boundary between
condensed and non-condensed ions.   

Fig.~\ref{fig:condIons_E_Sxxx}(c) shows that in weak fields, there are nearly
no coions condensed on the chain when $C_s<C_s^*$. In the high-salt region
above the equivalence point $C_s^*$, the coion condensation significantly
increases.  This is because the effective charge becomes positive, manifested
by the inversion of chain mobility as seen in Fig.~\ref{fig:mobility}(a), and
so attracts the coions to the chain.  After being constant in weak fields,
$N_{-1}^c$ decreases slightly with $E$ but then displays a large peak in strong
fields.  The  peak can be associated with the chain conformational transition
because tetravalent counterions condensed on a stretched, elongated chain are
more exposed to the bulk solution than ones condensed/wrapped in a unperturbed,
coiled or globular chain and hence, attract more coions.  The condensed coions
constitute an outer layer of the PE complex. Therefore, once the chain is fully
stretched, the increasing electric field can ``blow''  coions off due to the
weak condensation, causing $N_{-1}^c$ to decrease drastically.
  
\begin{figure}[htbp]
\begin{center}
\subfigure[]{\includegraphics[width=0.35\textwidth,angle=270]{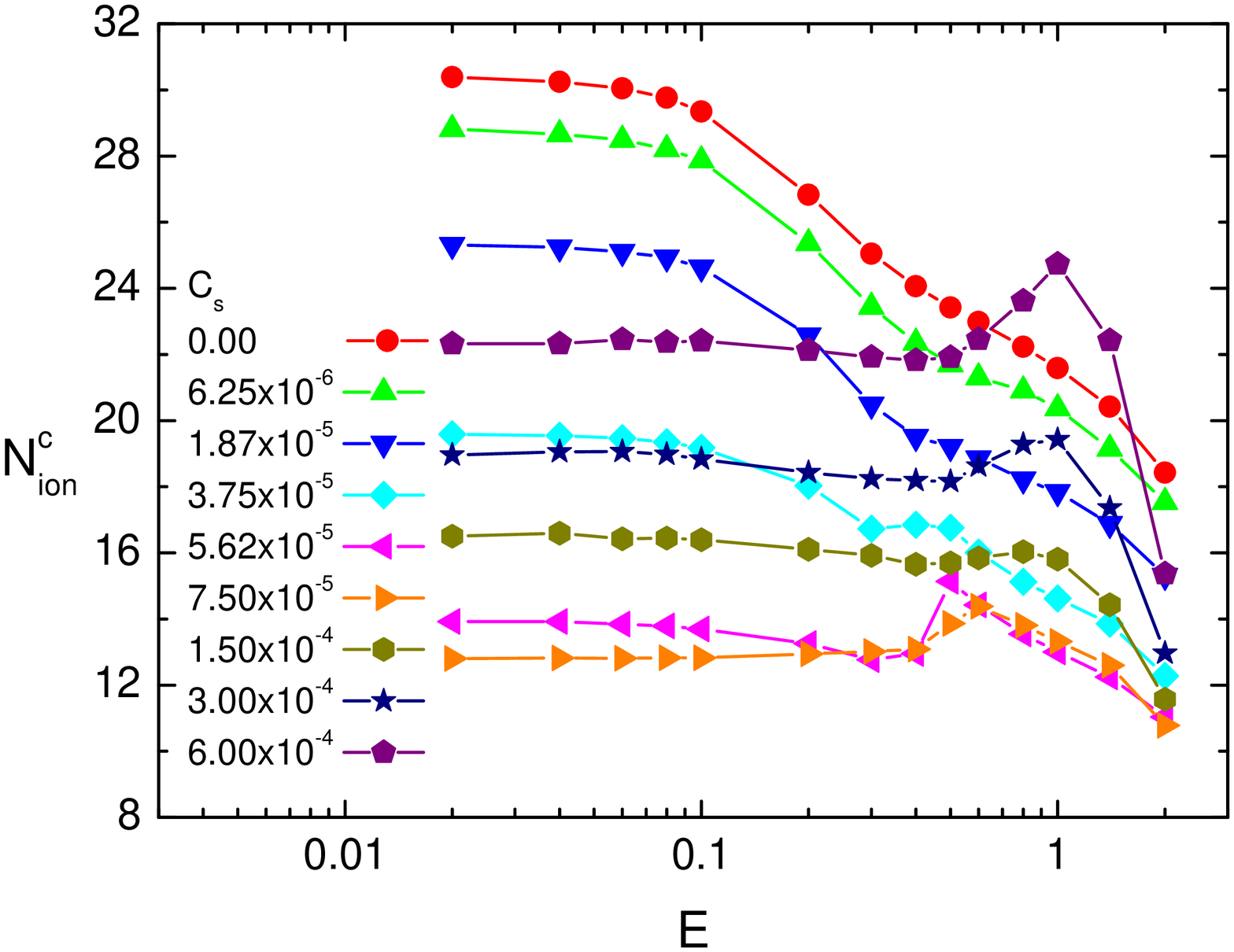}}
\subfigure[]{\includegraphics[width=0.35\textwidth,angle=270]{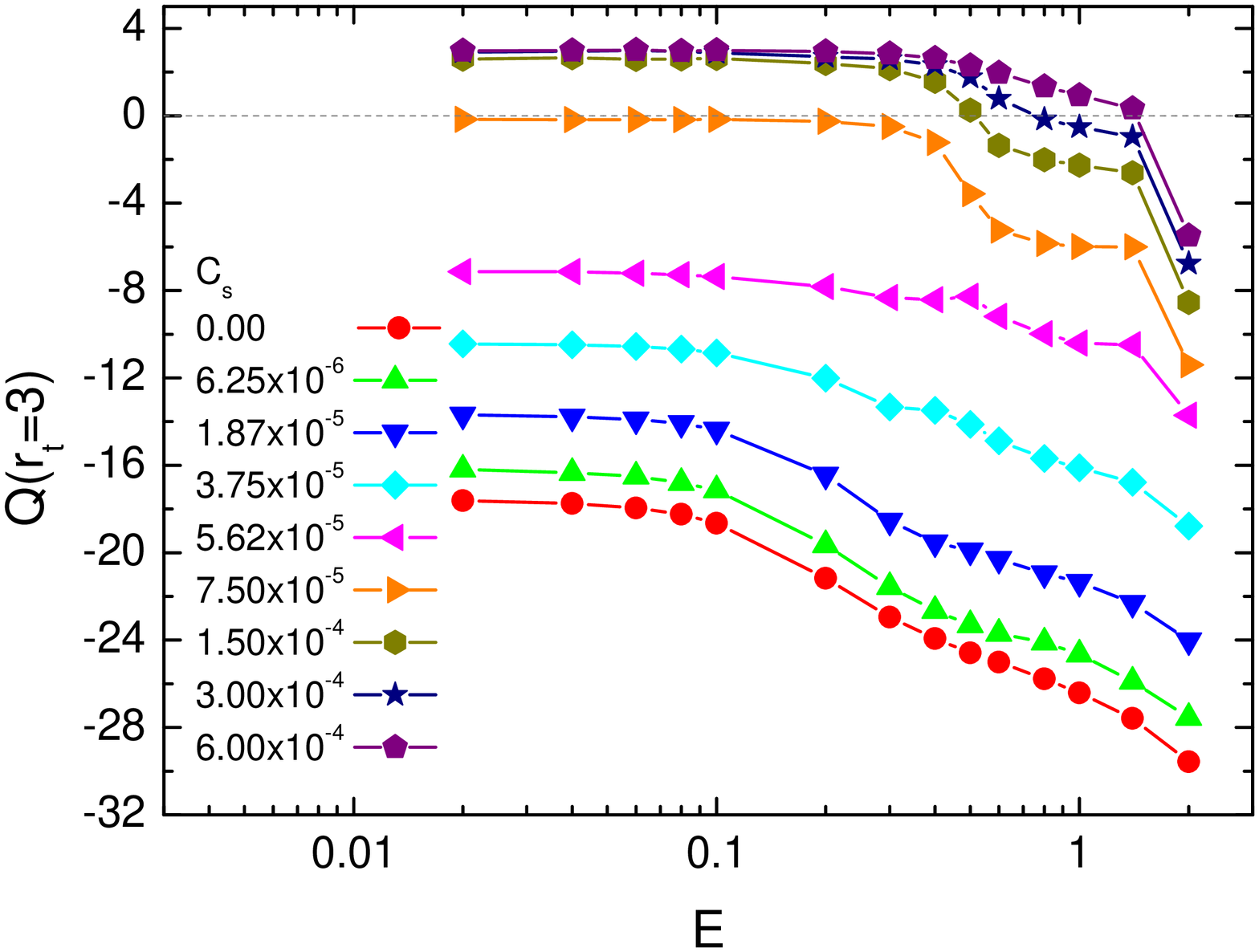}}
\caption{(a) Total number of condensed ions $N_{ion}^c$, and (b) total charge
of chain, $Q(r_t=\lambda_B)$, inside the condensation region $r_t=\lambda_B$,
as a function of $E$ at different $C_s$.} \label{fig:CtotCQc} \end{center}
\end{figure}

The above results show that the ions do not condense on the chain in a
universal way. Condensation depends strongly on the ion valency and the chain
conformation in electric fields.  The tetravalent counterions compete with the
monovalent counterions.  Therefore, while $N_{+4}^c$ increases with $C_s$,
$N_{+1}^c$ decreases.  On the other hand, the coions collaborate with the
tetravalent counterions; $N_{-1}^c$ increases with $C_s$ when $C_s>C_s^*$,
following the trend of $N_{+4}^c$.  The non-monotonic behavior of change can be
seen in Fig.~\ref{fig:CtotCQc}(a) where the total number of the condensed ions
$N_{ion}^c$ is plotted against $E$ at different $C_s$.
Nonetheless, the total charge inside the condensation region,
$Q(r_t=\lambda_B)$, (presented in Fig.~\ref{fig:CtotCQc}(b)) shows a more
regular, decreasing behavior.  We regarded $Q(r_t=\lambda_B)$ as the effective
charge and its variation compares well with the behavior of the chain mobility
$\mu_{pe}$ shown in Fig.~\ref{fig:mobility}(a).  By comparing the results of
Fig.~\ref{fig:CtotCQc}(a) and (b), we conclude that it is not the number of
ions but rather the total charge of the PE-ion complex that is constantly being
reduced by the strong electric field.

\subsection{Ion distribution along the chain and the mobilities}
As we have seen and studied, the PE-ion complex is polarized by an external
electric field. The polarization results from both the migration of the ions
condensed on the chain and the conformational change. In this subsection, we
study how the ion distribution along the chain varies with the electric field
and see how this distribution is related to the mobilities of the chain and the
condensed ions.  We associate a condensed ion to the monomer with which it is
closest, and calculated the mean number $n$ of the condensed ions associated
with each monomer.  The monomers are indexed by $i$ through the chain such that
the increment of the index follows the sense of the field direction and not
necessarily the position along the chain. \textcolor{\BLUE}{A front end and a
rear end of the chain are then defined to be the monomers with the smallest and
the largest index, respectively.}  Thus, $n(i)$ is the distribution function of
the condensed ions.  We study four representative cases of $C_s$.

\begin{figure}[htbp]
\begin{center}
\subfigure[]{\includegraphics[width=0.35\textwidth,angle=270]{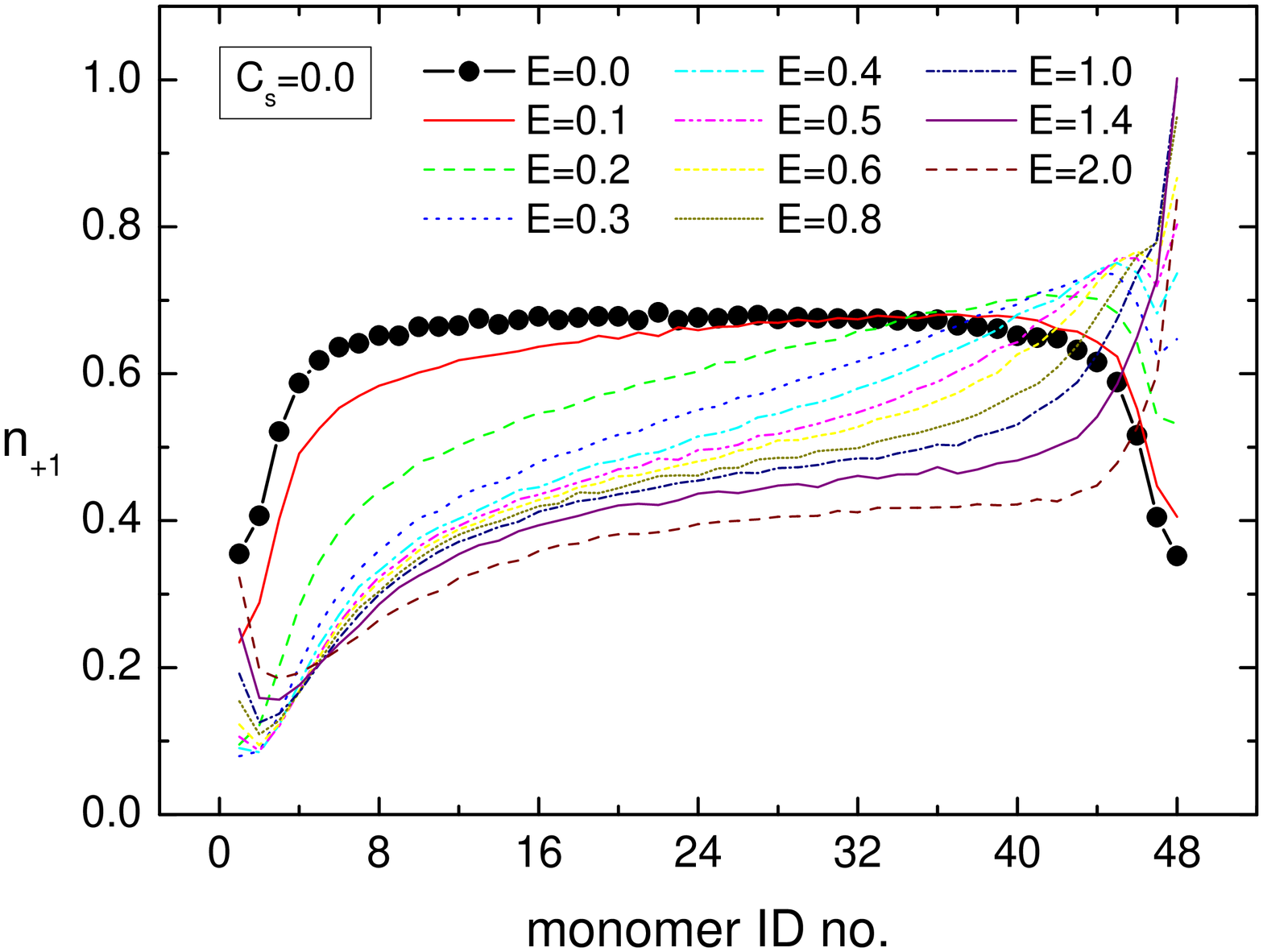}}
\subfigure[]{\includegraphics[width=0.35\textwidth,angle=270]{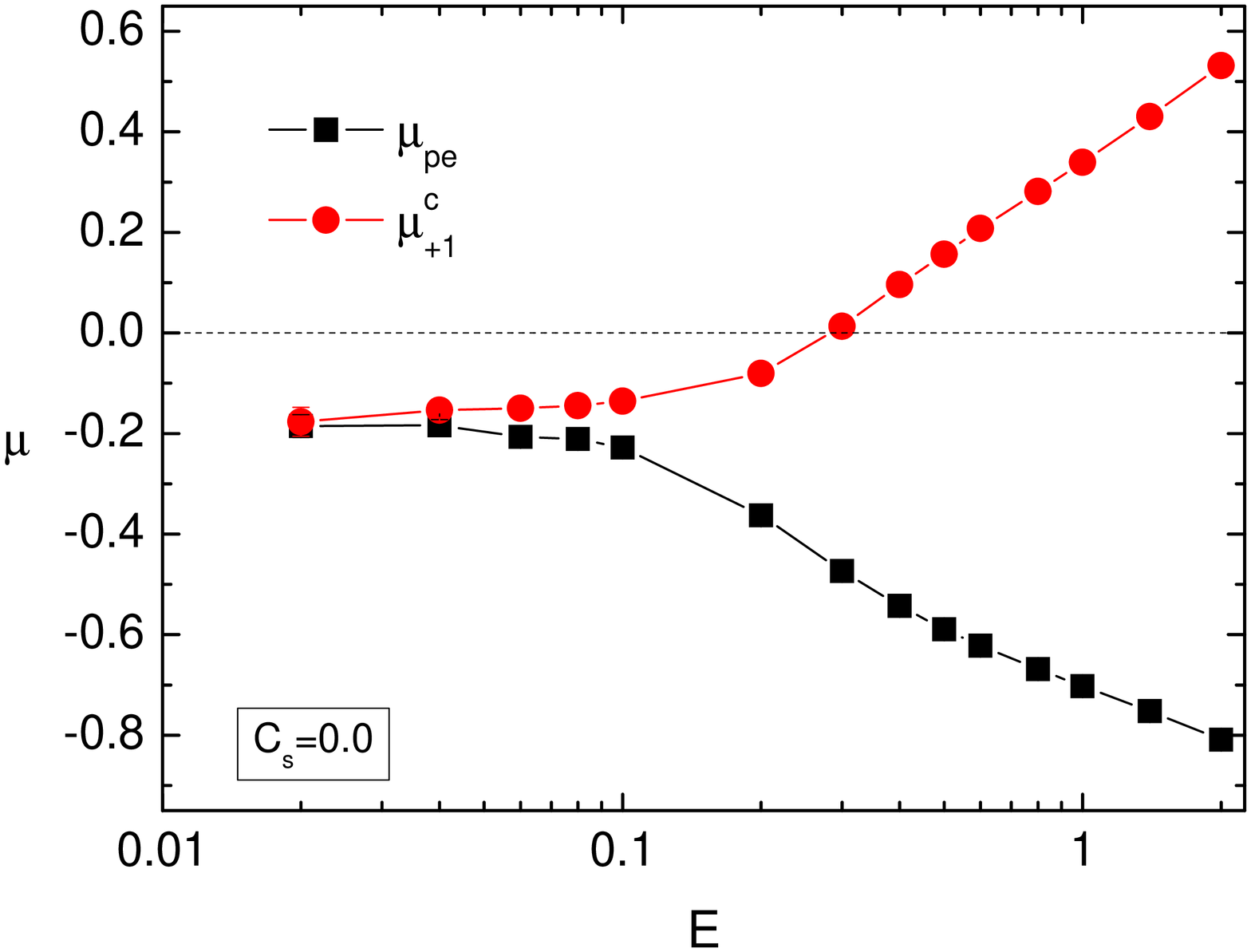}}
\caption{(a) Distribution of condensed monovalent counterion along chain,
$n_{+1}(i)$, at $C_s=0.0$ in different electric fields $E$.  The value of $E$
is indicated in the figure. (b) The mobility of the chain $\mu_{pe}$, and of
the condensed monovalent counterions $\mu_{+1}^c$, as a function of $E$.  }
\label{fig:distonPE_S0} \end{center}
\end{figure}

The first case considers the salt-free solution.  The only ions presented in
this case are the monovalent counterions. Fig.~\ref{fig:distonPE_S0}(a) shows
the number distribution of the condensed monovalent counterions on the chain,
$n_{+1}(i)$.

In the zero field limit, the $n_{+1}$ curve is flat in the interior region of
the chain and decreases near the two chain ends.  The decrease can be
attributed to the symmetry broken near the chain ends because there exists no
monomer outside the ends and hence the attractive force to condense the
counterions is weaker than in the interior. The whole distribution curve is
symmetric with respect to the middle of the chain.  The application of an
electric field breaks this symmetry by tilting the distribution curve against
the field direction, as shown.  The condensed monovalent counterions accumulate
near one end and are depleted near the other, leading to a polarization vector
pointing in the field direction.  The curve shifts downward as $E$ increases.
Since the area below the curve represents the number of ions condensed on the
chain, the downward-shifting shows the decrease of the number of condensed
counterions, which is consistent with the result in
Fig.~\ref{fig:condIons_E_Sxxx}(a). The shape of $n_{+1}$ evolves and becomes a
tangential curve when $E$ is very strong.  In order to understand the kinetics
of the condensed ions, we plot the mobility of the chain $\mu_{pe}$ and the
mobility of the condensed ions $\mu_{+1}^c$ in Fig.~\ref{fig:distonPE_S0}(b).
We found that $\mu_{+1}^c$ is basically equal to $\mu_{pe}$ in the weak fields,
which shows that the condensed counterions reside statically on the chain.
$\mu_{+1}^c$ deviates significantly from $\mu_{pe}$ when $E$ goes beyond $0.1$.
This deviation shows that there is relative motion between the condensed
monovalent counterions and the chain: The ions glide along the chain.  Since
the chain has finite length, the gliding means that the ions are dynamically,
rather than statically, condensed on the chain.  Once condensed onto the chain,
an ion glides along the chain backbone and accumulates at the chain end.  A
condensed ion at the chain end must then leave the chain to maintain a steady
number of the condensed ions.

\begin{figure}[htbp]
\begin{center}
\subfigure[]{\includegraphics[width=0.35\textwidth,angle=270]{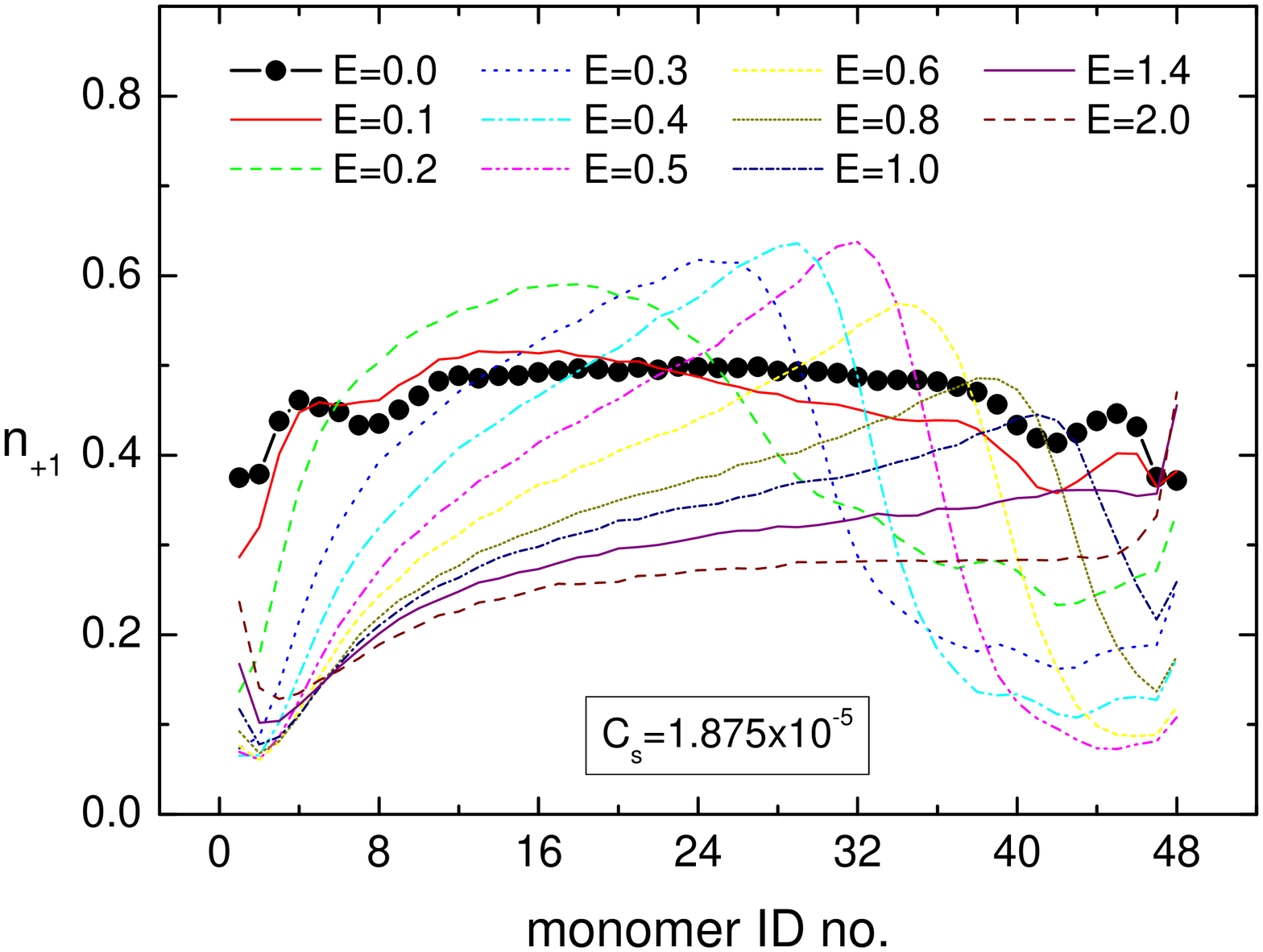}}
\subfigure[]{\includegraphics[width=0.35\textwidth,angle=270]{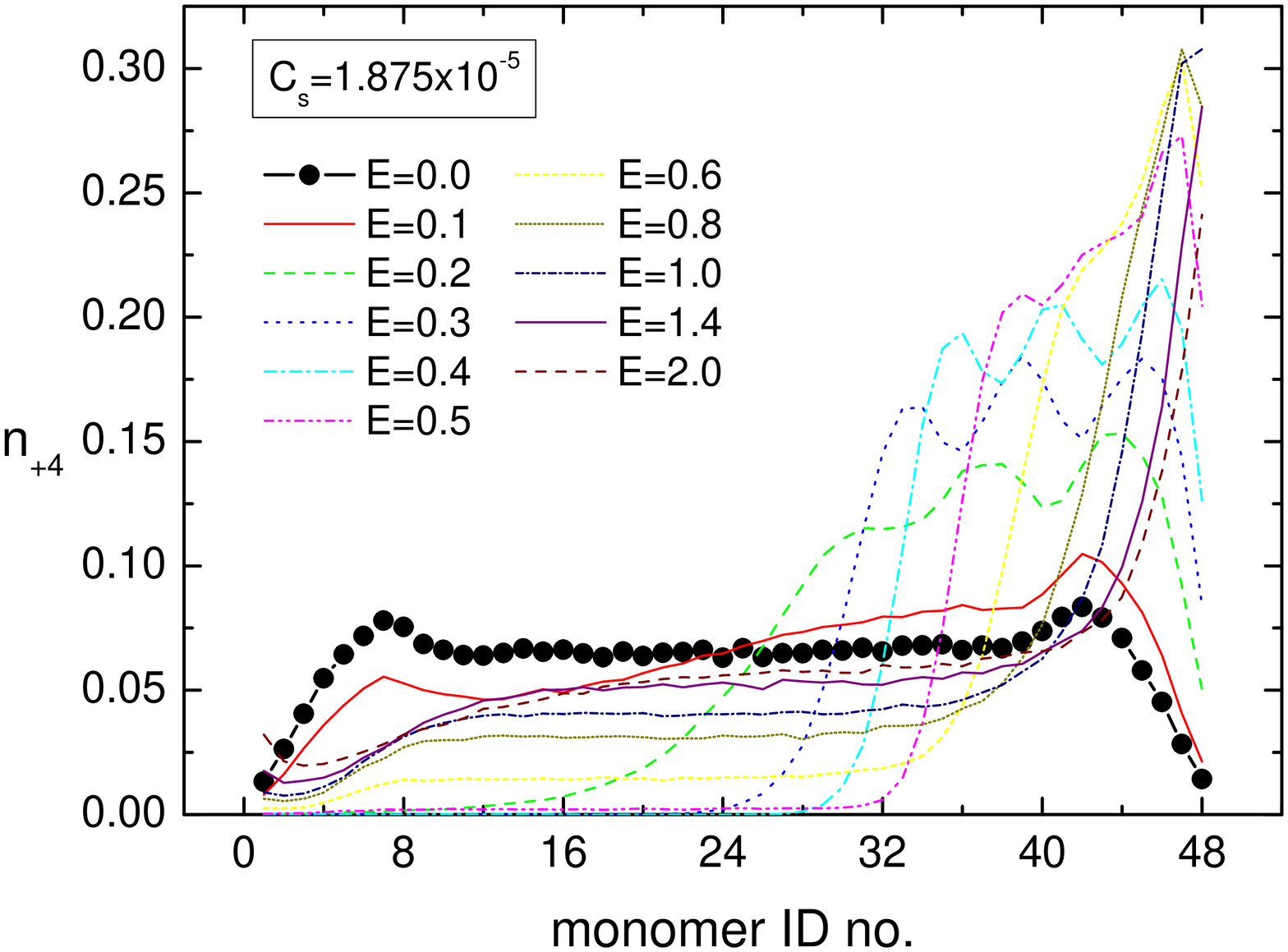}}
\subfigure[]{\includegraphics[width=0.35\textwidth,angle=270]{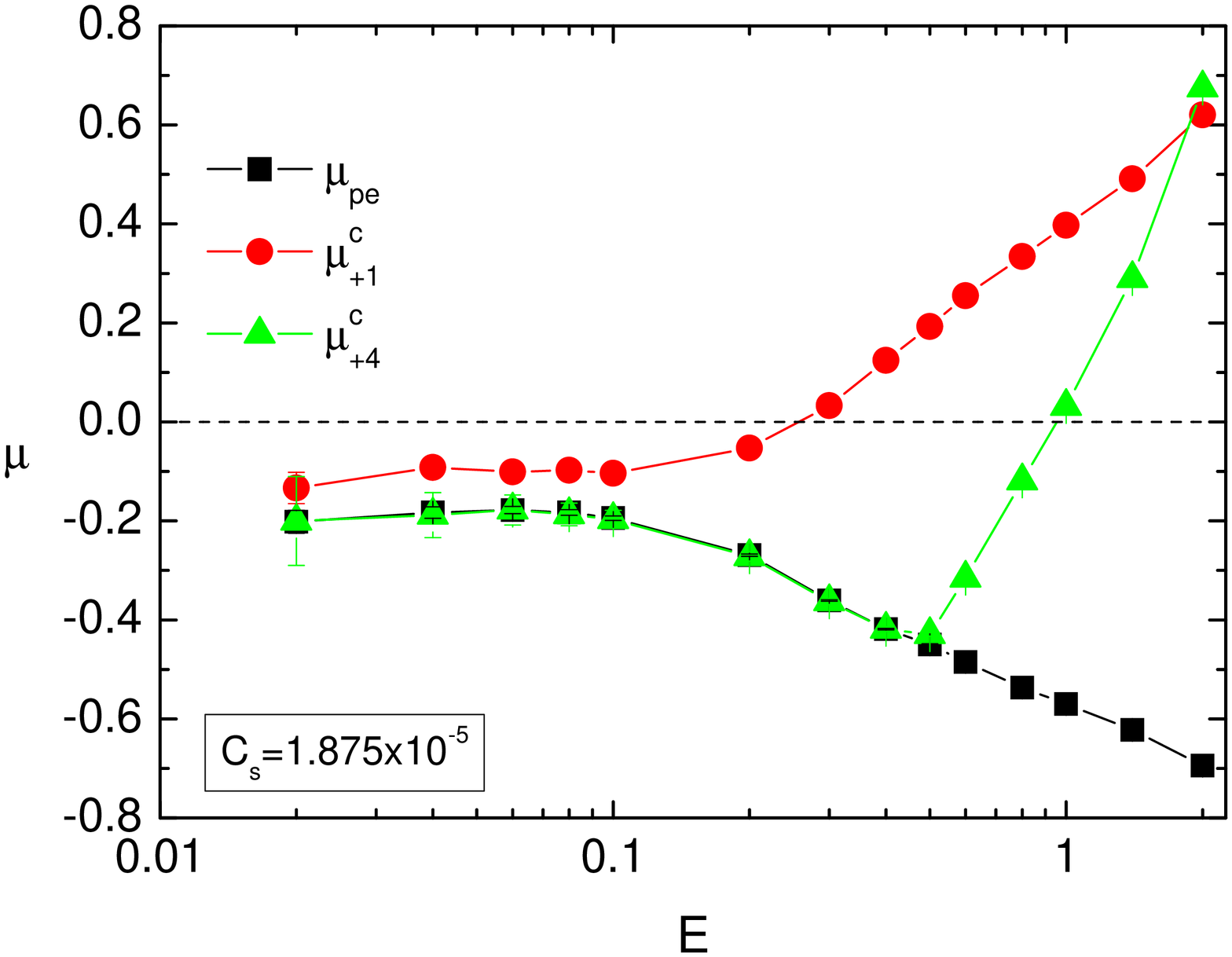}}
\caption{ (a) Distribution of condensed monovalent counterions on chain,
$n_{+1}(i)$, at $C_s=1.875\times 10^{-5} (<C_s^*)$  in different electric
fields $E$. The value of $E$ can be read in the figure. (b) Same as (a), but
for condensed tetravalent counterions,  $n_{+4}(i)$. (c) The mobility of the
chain $\mu_{pe}$, of the condensed monovalent counterions $\mu_{+1}^c$, and of
the condensed tetravalent counterions $\mu_{+4}^c$, as a function of $E$.  }
\label{fig:distonPE_S3} \end{center}
\end{figure}

For the second case, consider a salt concentration smaller than the equivalence
point, $C_s=1.875 \times 10^{-5}$. According to Fig.~\ref{fig:condIons_E_Sxxx},
there are nearly no coions condensing onto the chain. Therefore, we show only
the distribution functions for the condensed monovalent counterions,
$n_{+1}(i)$, and for the condensed tetravalent counterions, $n_{+4}(i)$. The
results are presented in Figs.~\ref{fig:distonPE_S3}(a) and (b), respectively.

Observe that in the zero field limit, the $n_{+1}(i)$ and $n_{+4}(i)$ curves
are both flat in the interior and symmetric to the middle of the chain.  When
an electric field is applied, the symmetry is broken.  $n_{+1}$ displays a
large skewed peak in the interior region of the chain, while $n_{+4}$ shows
multiple peaks close to the rear end. The number of peaks in $n_{+4}$ is 3 for
$0.2 \le E \le 0.5$, which corresponds exactly to the number of the tetravalent
counterions added at this salt concentration, \textit{i.e.}, the tetravalent
counterions completely condense.  The distinct peaks reveals that the ions are
localized on the chain, similar to a 1-dimensional crystal. This localization
happens when appropriate strength of electric field is applied.  If $E$ is
weak, the ions cannot be localized and the $n_{+4}$ curve simply tilts.  If $E$
is strong, they are pushed forcibly to the rear end and only one peak is
visible.  Similar phenomena have been observed for other $C_s<C_s^*$.  The peak
of the $n_{+1}$ curve appears right after the peaks of the $n_{+4}$ curve.
This non-overlapping of the two species results from mutual exclusion between
the condensed monovalent counterions and tetravalent counterions due to their
Coulomb repulsion. Since the external electric force exerted on an ion is
proportional to the ion valency, the tetravalent counterions are pushed more
strongly by the electric fields and move closer to the chain end. The
monovalent counterions suffer a weaker force and condense just behind the
tetravalent counterions.  

We also plot the mobility of the chain and the condensed ions as a function of
$E$ in Fig.~\ref{fig:distonPE_S3}(c) to study the kinetics of the system.  We
can see that the mobility of the tetravalent counterions $\mu_{+4}^c$ coincides
with $\mu_{pe}$ in small fields and deviates from it when $E>0.5$.  The tight
coincidence shows that the tetravalent counterions are tightly bound to the
chain and move with it. The mobility of the condensed monovalent counterions
$\mu_{+1}^c$ also takes a value close to $\mu_{pe}$ when $E\le 0.1$ but not as
close as when $C_s=0.0$ (cf.~Fig.~\ref{fig:distonPE_S0}.)  The tetravalent
counterions repel the condensed monovalent counterions on the chain, which
loosens the condensation.  \textcolor{\BLUE}{Recall that the deviation of
$\mu_{+4}^c$ from $\mu_{pe}$ occurs around $E=0.5$. This is about 4 times the
field strength as when $\mu_{+1}^c$ begins to deviate from $\mu_{pe}$.  And the
number 4 is equal to the valency ratio between the two species of the
counterions,  which can be explained by law of friction.  Assume that the
friction coefficient of the surface is effectively $\xi$.  The criterion for a
condensed ion to glide along the chain surface is governed by $F_E \ge \xi F_n
$, where $F_E=ZeE$, and $F_n$ is the force normal to the chain surface, which
is estimated by $k_BT \lambda_B Z^2e^2/\sigma^2$.  The square of $Z$ appears in
$F_n$ because of the matching (or interaction) of a $Z$-valent counterion with
$Z$ monomers when it condenses on the chain.  Therefore, the threshold electric
field to glide a condensed counterion is approximately proportional to the
valency.} 
 
\begin{figure}[htbp]
\begin{center}
\subfigure[]{\includegraphics[width=0.35\textwidth,angle=270]{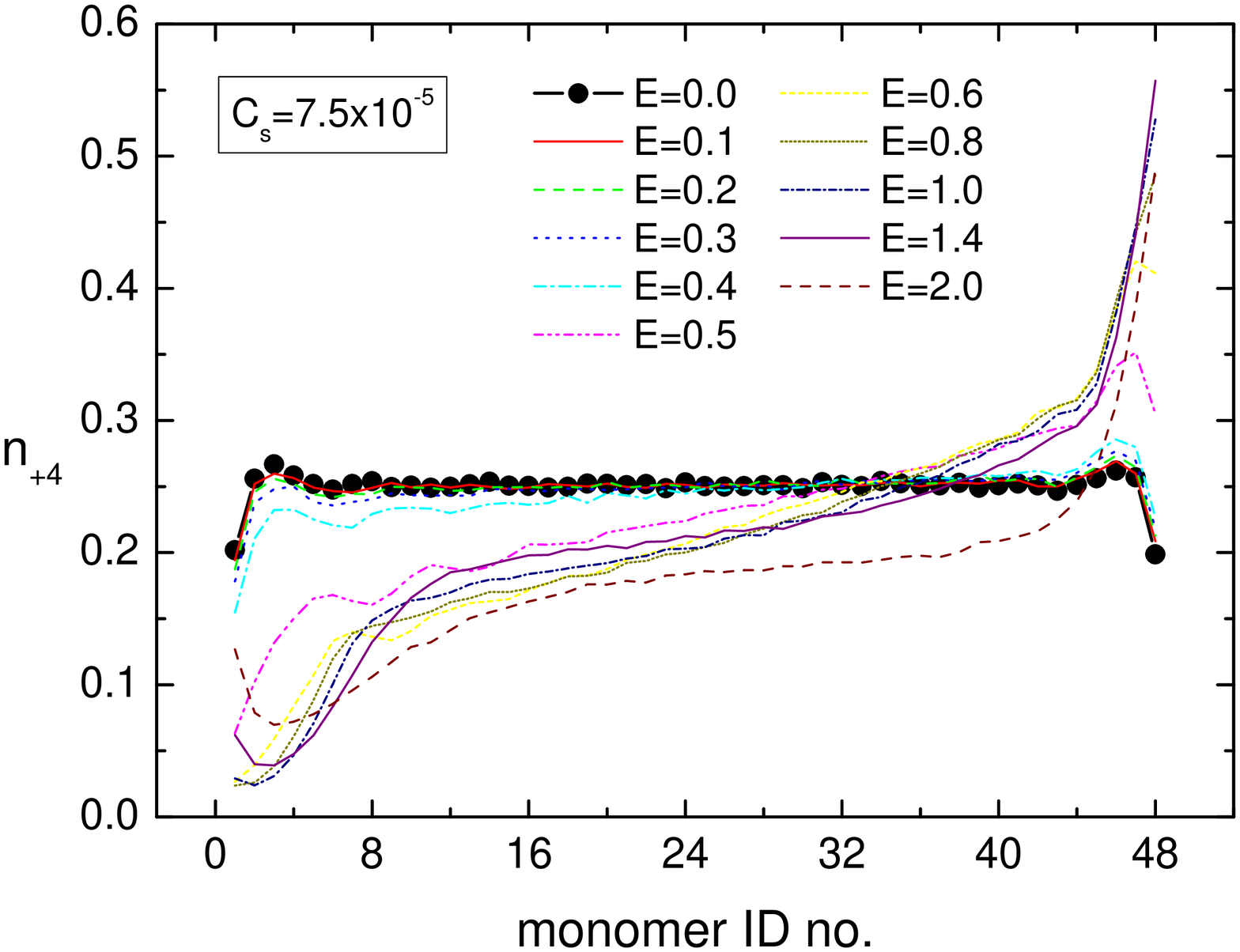}}
\subfigure[]{\includegraphics[width=0.35\textwidth,angle=270]{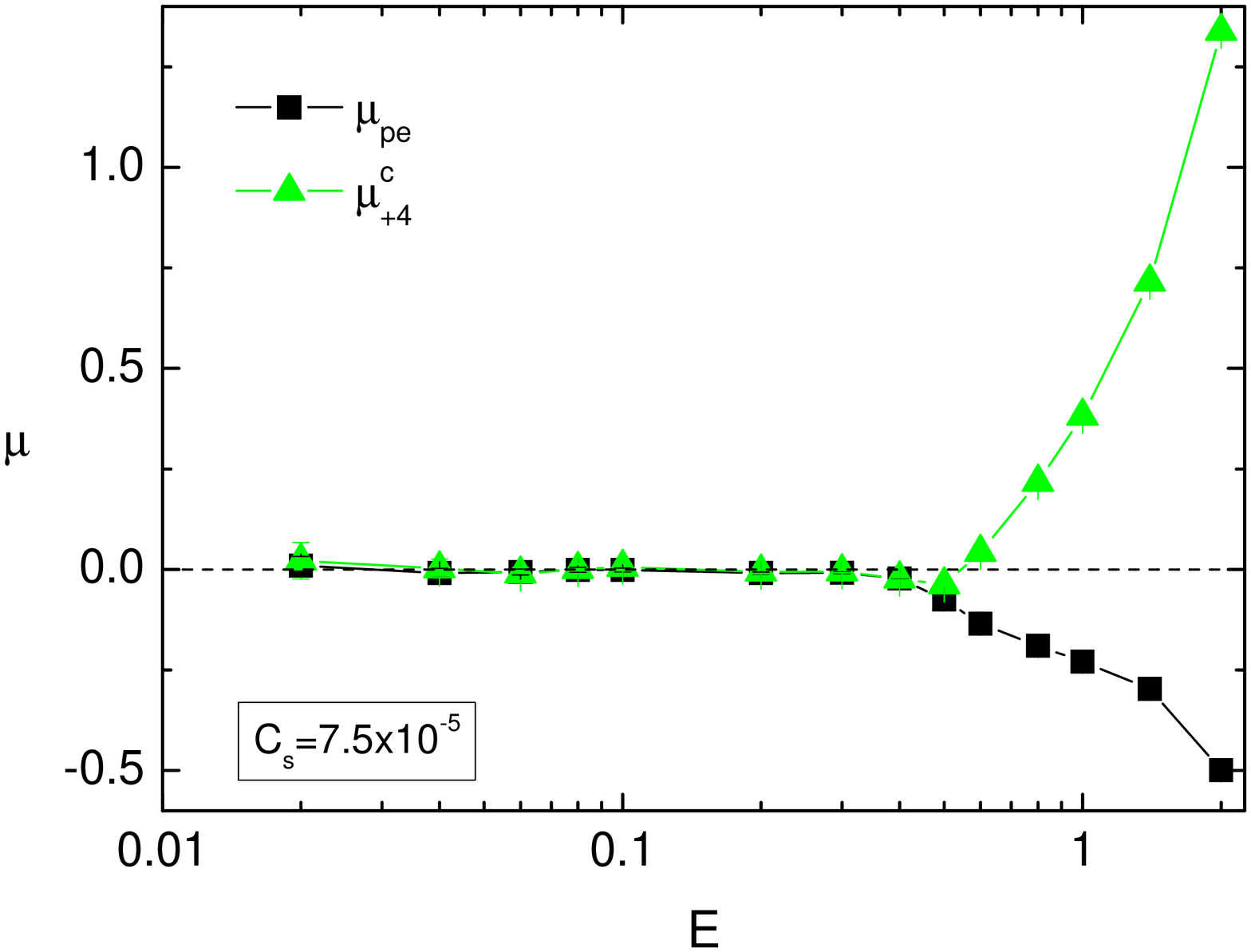}}
\caption{(a) Distribution of condensed tetravalent counterions on chain at
$C_s=7.5\times 10^{-5}(=C_s^*)$ in different electric fields $E$. The value of
$E$ can be read in the figure. (b) The mobility of the chain $\mu_{pe}$ and of
the condensed tetravalent counterions $\mu_{+4}^c$ as a function of $E$. }
\label{fig:distonPE_S12} \end{center}
\end{figure}

The third case studies the system at the equivalence point $C_s^*$.  At this
salt concentration, no monovalent counterions or coions condense on the chain;
so only the distribution function for the tetravalent counterions is shown.
The results of $n_{+4}(i)$ and the mobilities for the chain and the ions are
presented as a function of $E$ in Fig.~\ref{fig:distonPE_S12}.

The behavior of $n_{+4}(i)$ looks similar to  $n_{+1}(i)$ in
Fig.~\ref{fig:distonPE_S0}(a).  The curve tilts, shifts downward, and
eventually shows a tangential distribution as $E$ increases.  Since the
condensed ions have large valency, the condensation is strong.  The chain is
charge-neutralized by the condensed ions and Fig.~\ref{fig:distonPE_S12}(b)
shows that $\mu_{pe}$ is zero when $E<0.5$.  Above $E>0.5$, the electric force
exerted on the condensed ions can overcome the friction force between the ions
and the chain surface; hence there is relative motion between the ions and the
chain.

\begin{figure}[htbp]
\begin{center}
\subfigure[]{\includegraphics[width=0.35\textwidth,angle=270]{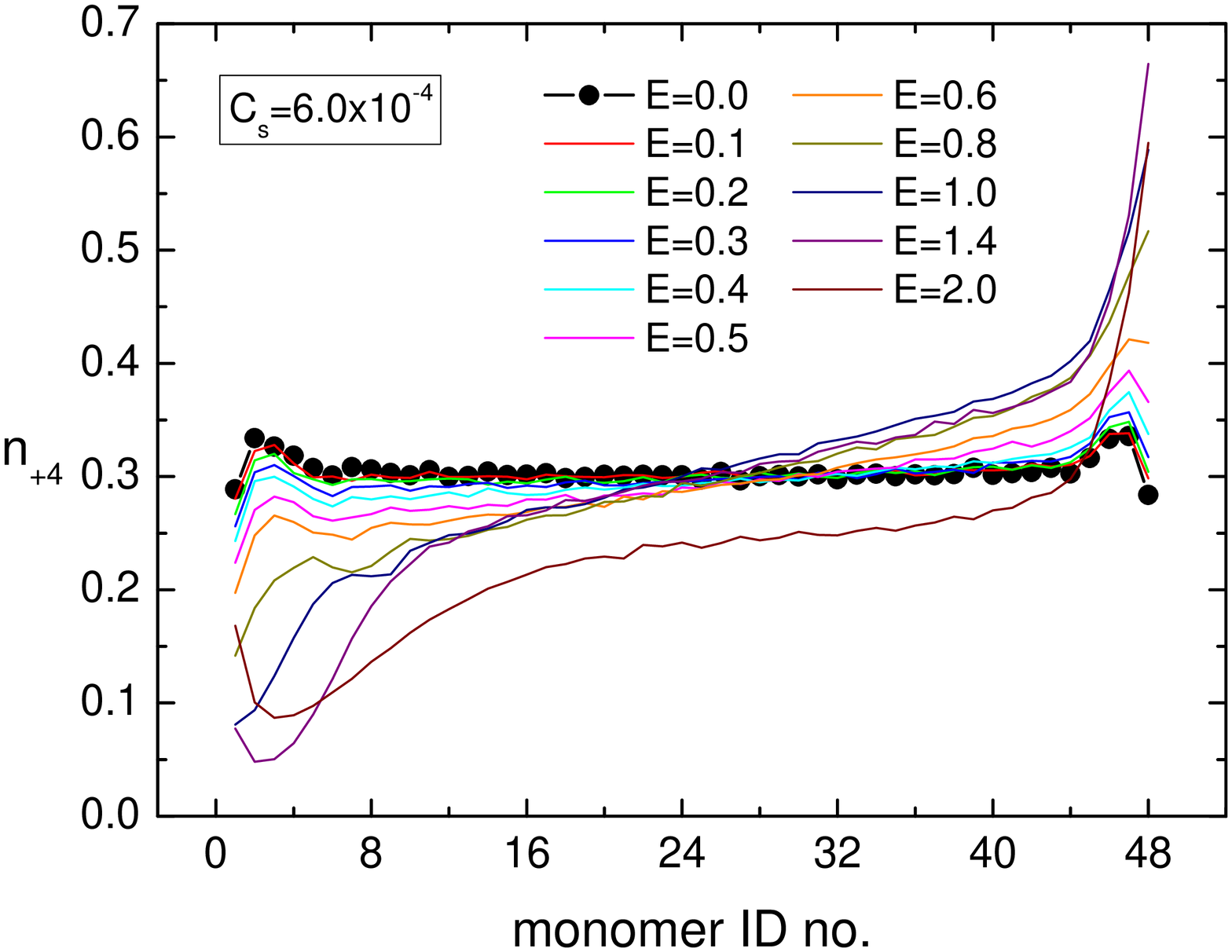}}
\subfigure[]{\includegraphics[width=0.35\textwidth,angle=270]{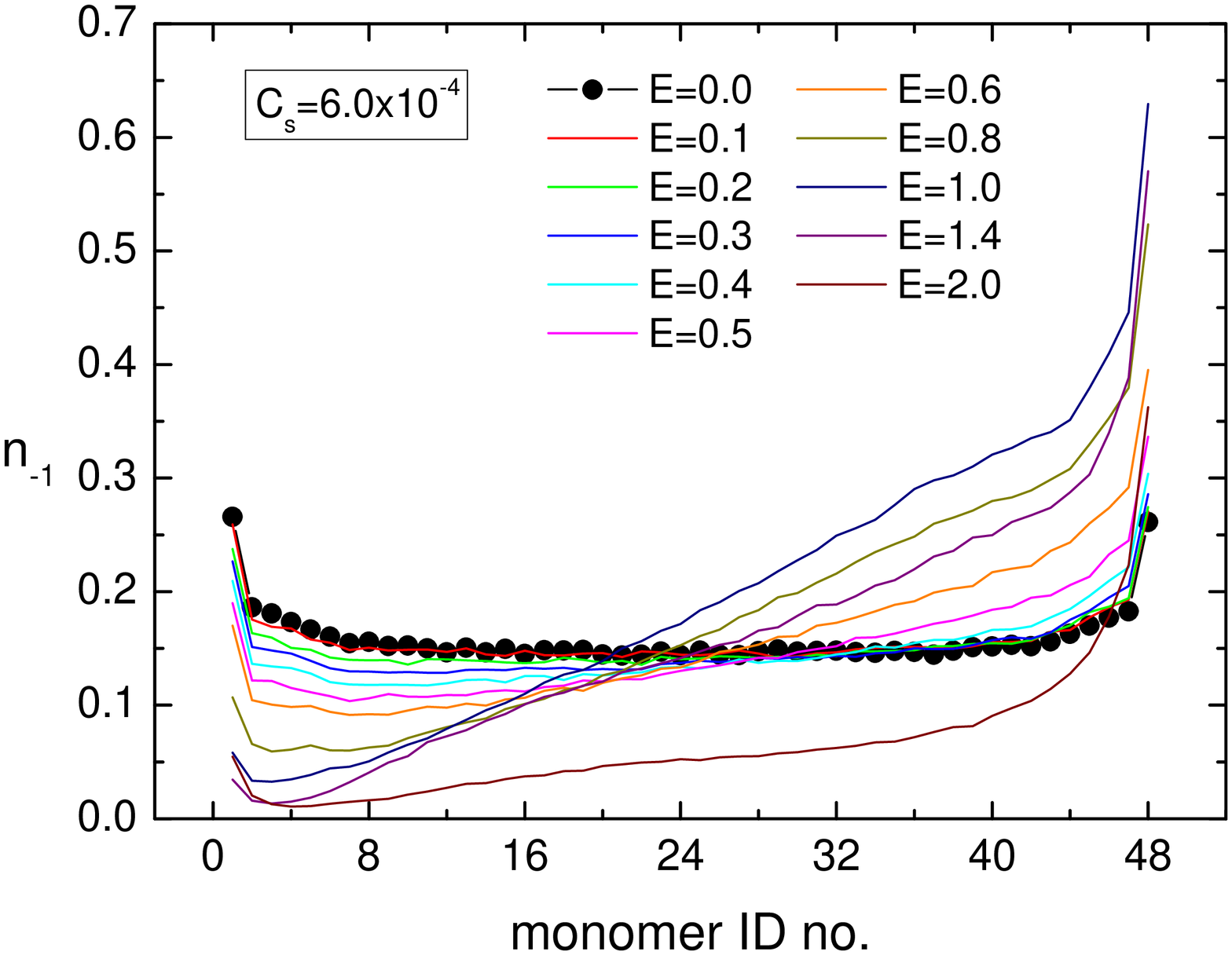}}
\subfigure[]{\includegraphics[width=0.35\textwidth,angle=270]{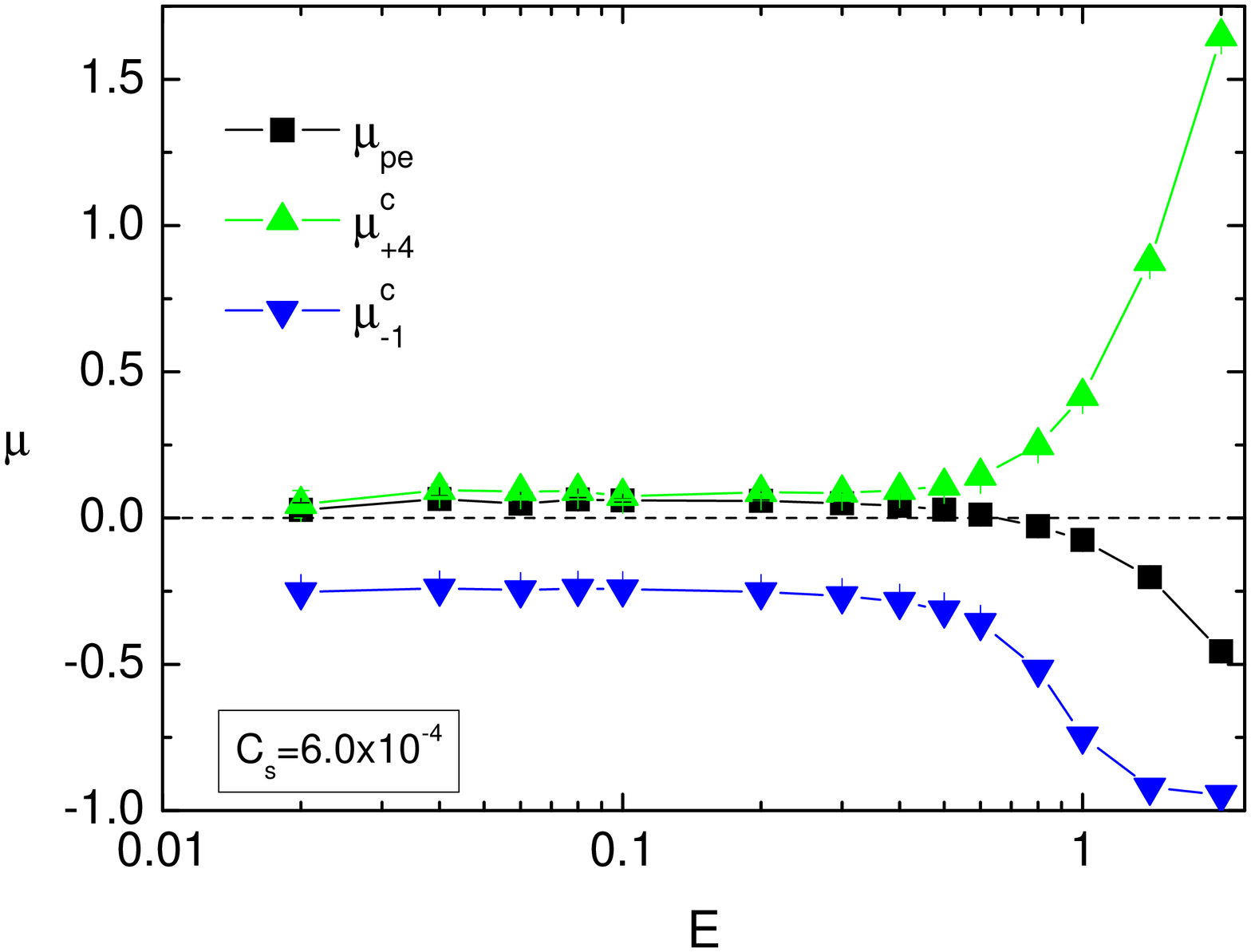}}
\caption{ (a) Distribution of condensed tetravalent counterions on chain,
$n_{+4}(i)$,  at $C_s=6.0\times 10^{-4} (>C_s^*)$ in different electric fields
$E$. The value of $E$ can be read in the figure. (b) Same as (a), but for
condensed coions $n_{-1}(i)$. (c) The mobility of the chain $\mu_{pe}$, of the
condensed tetravalent counterions $\mu_{+4}^c$, and of the condensed coions
$\mu_{-1}^c$, as a function of $E$. } \label{fig:distonPE_S96} \end{center}
\end{figure}

In the fourth case, we study a salt concentration higher than the equivalence
point.  Figs.~\ref{fig:distonPE_S96}(a) and \ref{fig:distonPE_S96}(b) present,
respectively, the distributions of the condensed tetravalent counterions
$n_{+4}(i)$ and of the coions $n_{-1}(i)$ on the chain at $C_s=6.0 \times
10^{-4}$.  The distribution of the monovalent counterion $n_{+1}(i)$ is not
shown because very few of them condense onto the chain.

The distribution $n_{+4}(i)$ is similar to the one at $C_s^*$ but the curve is
shifted upward by a value of about $0.5$. This upward-shifting shows that the
number of the condensed tetravalent counterions exceeds the number needed to
neutralize the chain. The chain is overcharged, and hence attracts coions onto
it. The distribution $n_{-1}(i)$ tilts when $E$ is applied. The tilting is not
toward the front end but rather toward the rear end and looks similar to
$n_{+4}(i)$. It suggests that the condensation of the coions are, in fact,
mediated by the condensed tetravalent counterions.  The condensation of coion
is weak because the mobility for the condensed coions $\mu_{-1}^c$ is not zero
even for the small set $E$ fields simulated, as plotted in
Fig.~\ref{fig:distonPE_S96}(c).  The negative value of $\mu_{-1}^c$ tells us
that the condensed coions migrate constantly toward the chain front end.
Surprisingly, the number of coions is less elevated at this end.  This suggests
that the chemical potential of the coion condensation is higher at this end.
The tetravalent counterions condense more numerously at the rear of the chain,
lowering the chemical potential there.  The mobility $\mu_{+4}^c$ is identical
to $\mu_{pe}^c$ when $E<0.5$, which once again shows the strong condensation of
tetravalent counterions.  The tetravalent ions reside on the chain and move
along with it.  The positive value of the mobilities demonstrates that the
effective chain charge becomes positive due to the over-condensation of
tetravalent counterions.  When $E$ is strong enough to overcome the friction on
the chain surface, the kinetics for the condensed ions and the chain are
decoupled.  $\mu_{+4}^c$ increases but $\mu_{pe}^c$ decreases.  The magnitudes
of the mobilities both increase.  $\mu_{-1}^c$ asymptotically approaches -1,
the value expected for a coion drifting in the bulk solution. 

\textcolor{\RED}{We remark that the distributions of condensed ions studied here
might be understood in the framework of sedimentation. In a dilute solution,
sedimentation theory describes a barometric density profile $\rho(z)=\rho_0
\exp(-z/L)$ for neutral particles, where $z$ is the position in the
gravitational direction and $L=k_BT/(mg)$ is the characteristic length (also
equal to the mean height of sediments) with $m$ being the particle buoyant mass
and $g$ the gravitational acceleration. For charged particles, the mean height
is shown to extend to $ZL$ owing to electrostatic interactions~\cite{vanRoij03,
cuetos06}. In our work, the condensed ions are under the action of an electric
field, which can be analogously regarded as sedimenting in a gravitational
field with the force $mg$ effectively replaced by the net force $ZeE-\xi F_n$
acting on a condensed ion.  This analogy makes sense only when $E$ is strong
enough to displace the condensed ions and totally unfold the chain, so that the
index $i$ is linearly mapped to $z$.  Following the index number in a reverse
order, we observe that $n_{+1}(i)$ in Fig.~\ref{fig:distonPE_S0} displays as an
exponential function for $E>0.5$, except near the two chain ends where some
edge effect appears. The characteristic length $L$ of the exponential can be
seen to decrease with increasing $E$, which follows the depiction of the
sedimentation theory. Similar results are found for $n_{+4}(i)$ in
Figs.~\ref{fig:distonPE_S12} and \ref{fig:distonPE_S96} when $E$ is strong.  In
the second case studied (in Fig.~\ref{fig:distonPE_S3}), we have two species of
counterions condensing on the chain. Zwanikken and van Roij have developed a
mean-field theory for the sedimentation of multicomponent charged colloids,
based upon a Poisson-Boltzmann approach~\cite{zwanikken05} A segregation of
layering charged colloids of valency $Z_i$ was demonstrated, in which the order
of the sediments, from the bottom to top, follow the increasing order of the
product $Z_iL_i$~\cite{zwanikken05, cuetos06}.  Since $L_i=k_BT/(Z_i e E-\xi
F_n)$ and $F_n$ is approximately proportional to $Z_i^2$, the tetravalent
counterions have a larger $Z_iL_i$ than the monovalent ones. The theory thus
predicts a profile along the chain which has the inverse order of what we
obtained in the simulations. This difference arises because some mechanism
cannot occur in mapping to a sedimentation system, for example, the dynamic
condensation of counterions on a chain.  The condensed monovalent counterions
are stripped off in an electric field, more easily than the tetravalent ones.
Consequently, the tetravalent counterions stay near the rear end of chain with
longer time and the monovalent ones coming from the front are then stuck behind
the tetravalent, which produces the observed profile.  }

\subsection{Dependence of chain mobility on chain length}
\begin{figure}[htbp]
\begin{center}
\subfigure[]{\includegraphics[width=0.4\textwidth,angle=270]{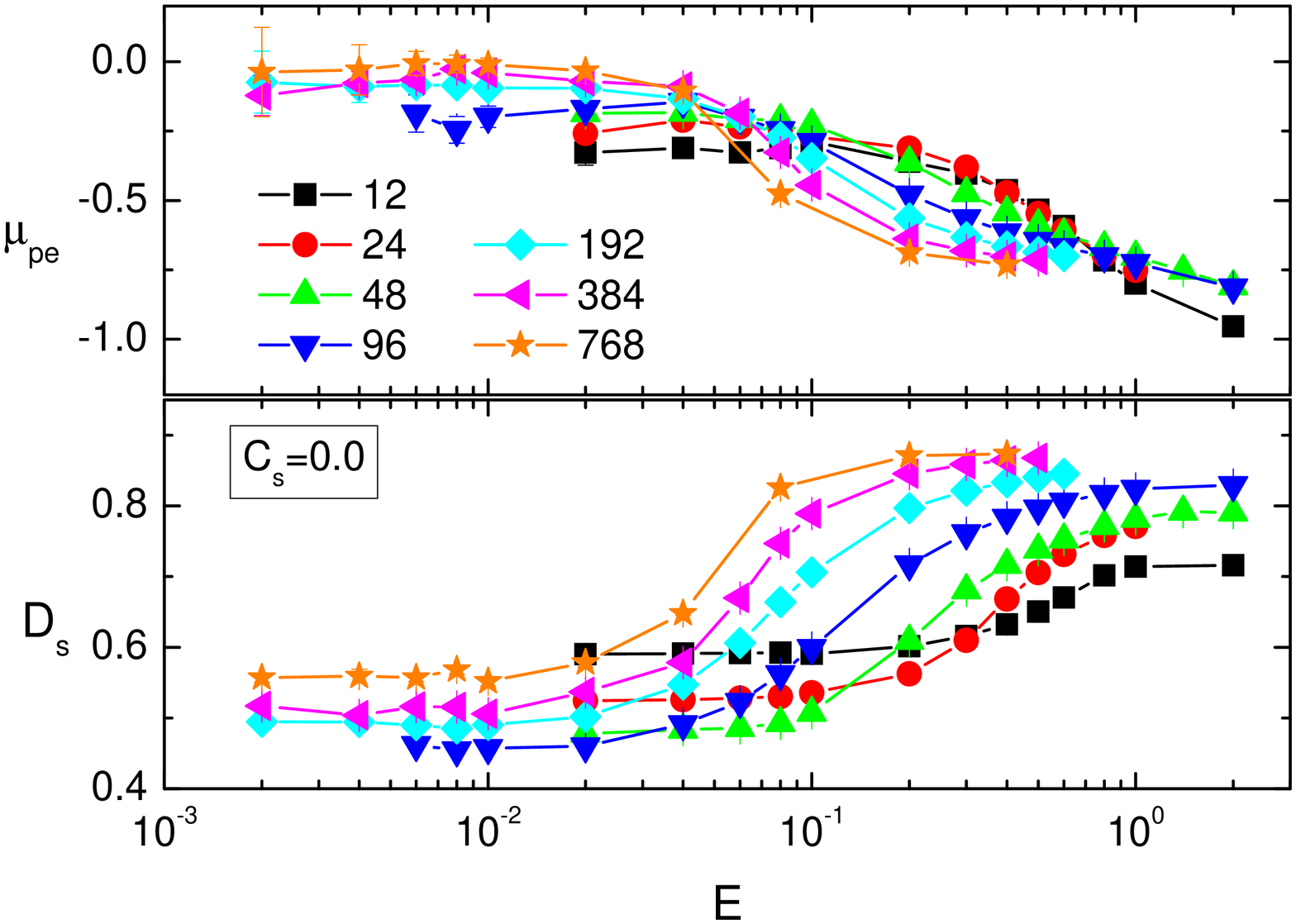}}
\subfigure[]{\includegraphics[width=0.4\textwidth,angle=270]{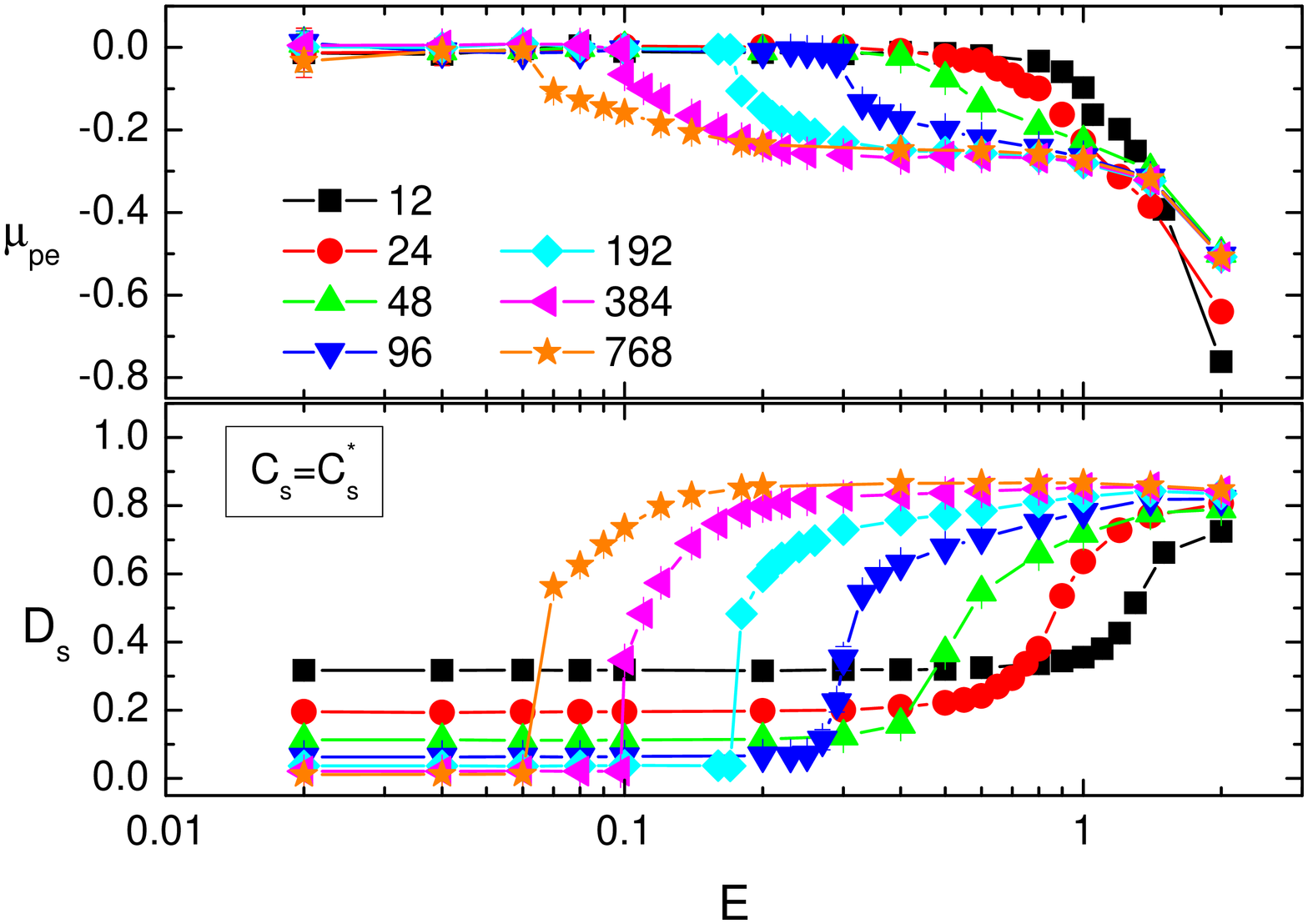}}
\caption{Chain mobility $\mu_{pe}$ and degree of chain unfolding $D_s$ versus
$E$ (a) in salt-free solutions and (b) in solutions at equivalence point
$C_s^*$. Each curve denotes the results run at one chain length $N$ and the
value of $N$ is indicated in the upper panel.} \label{fig:P1Nxx_mobility_rc3_Sxx}
\end{center}
\end{figure}

Finally we studied the variation of chain mobility $\mu_{pe}$ with chain length
$N$, as the applied electric $E$ increases.  The results are presented in
Fig.~\ref{fig:P1Nxx_mobility_rc3_Sxx}, panel (a) and (b), at two salt
concentrations, $C_s=0.0$ and $C_s^*$, respectively, for $N$ varying from 12 to
768. We have seen in Sec.~III-C that the magnitude of $\mu_{pe}$ largely 
increases when a chain unfolds.

Observe that the longer the chain length, the weaker the field strength that is
needed to unfold the chain.  Accompanying the unfolding, $|\mu_{pe}|$ increases
and the mobility reaches a constant value in the field region between $E=0.1$
and $1$.  For a stronger $E$, the magnitude of $\mu_{pe}$ increases again. It
is because the condensed ions can be further stripped off the chain,  which
renders the effective chain charge more negative, and thus, the chain drifts
faster. The electric field required to change the chain mobility
depends on the chain length.  It provides a unique mechanism to
electrophoretically separate PEs by $N$.  Especially at $C_s=C_s^*$, the chains
are initially neutralized by the multivalent counterions. They do not drift in
weak fields.  When an appropriate electric field is applied, the longer chains
unfold and gain mobility, and as a consequence, drift away and are separated
from the shorter chains.

\begin{figure}[htbp]
\begin{center}
\includegraphics[width=0.4\textwidth,angle=270]{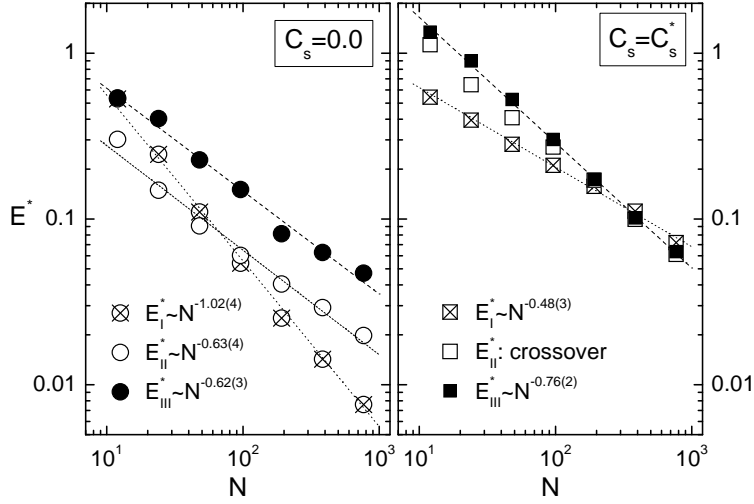}
\caption{\textcolor{\BLUE}{Critical electric field $E^*$ as a function of chain
length $N$ for $C_s=0.0$ and $C_s=C_s^*$. $E^*$ is determined by the three
ways: (1) the polarization energy equal to $k_BT$, (2) the onset point of
$D_s$, and (3) the inflection point of $D_s$. The three estimators are denoted
by $E^*_{I}$, $E^*_{II}$, and $E^*_{III}$, respectively.} }
\label{fig:Einfl_N12N768} \end{center}
\end{figure}

Netz predicted that the unfolding electric field $E^*$ should scale as
$N^{-3\nu/2}$ where $\nu$ is the swelling exponent of chain size.  We found in
Fig.~\ref{fig:P1Nxx_mobility_rc3_Sxx} that the chain unfolding, and also the
mobility change, are not sharp transitions in electric fields.  Consequently,
it is not easy to determine the onset electric field for chain unfolding,
$E^*_{II}$, with a good accuracy. Therefore, to verify the Netz' prediction, we
calculated, in addition, the electric field $E^*_{III}$, which is the
inflection point on the $D_s$-curve. \textcolor{\BLUE}{The results are plotted
in Fig.~\ref{fig:Einfl_N12N768} as a function of $N$, together with $E^*_{I}$
and $E^*_{II}$.}

\textcolor{\BLUE}{The three data sets at $C_s=0.0$ lie on straight lines in the
log-log plot: The critical electric field does indeed follow a power-law
function.  Linear least-square fit yields that $E^*_{I}$ scales as
$N^{-1.02(4)}$, $E^*_{II}$ as $N^{-0.63(4)}$, and $E^*_{III}$ as
$N^{-0.62(3)}$.  The scaling law for $E^*_{I}$ is a refinement of our previous
work~\cite{hsiao08b}. It behaves differently to $E^*_{II}$, which is calculated
directly from the chain size variation, and thus fails to predict the threshold
field to unfold a chain. The failure arises from oversimplified setting of the
critical polarization energy for chain unfolding $W_{pol}^*$ to the thermal
energy $k_BT$, which should be also a function of chain length and salt
concentration, as shown in Fig.~\ref{fig:P1Nxx_PxE_rc3}. The two fitting lines
for $E^*_{II}$ and $E^*_{III}$ run parallel to each other. Consequently, the
ratio of $E^*_{III}$ to $E^*_{II}$ is a constant, suggesting that either
measure of $E^*$ is acceptable.  At $C_s=C_s^*$, $E^*_{I}$ scales as
$N^{-0.48(3)}$, a wrong prediction for the unfolding field.  For
$E^*_{II}$ and $E^*_{III}$, markedly different behavior is observed.  Only the
$E^*_{III}$ data follow a scaling law with the exponent equal to $-0.76(2)$.
The onset field point $E^*_{II}$ exhibits a crossover from an exponent close to
$-0.63$ at small polymerizations to $-0.76$ when $N>100$ which agrees with the
$E^*_{II}$ value.  This crossover reflects the fact that the transition of
chain unfolding becomes sharper when chain length increases, as seen in
Fig.~\ref{fig:P1Nxx_mobility_rc3_Sxx}(b). } 

To check Netz' theory~\cite{netz03a, netz03b}, we calculate the swelling
exponent $\nu$ of chains in the zero field limit through the scaling relation
$R_g^2\sim N^{2\nu}$.  Our simulations show that $\nu=0.92(1)$ at $C_s=0.0$ and
$\nu=0.33(1)$ at $C_s=C_s^*$.  Therefore, according to Netz, the critical field
should scale as $N^{-1.38(2)}$ and $N^{-0.50(2)}$, respectively. However,  the
scaling law obtained in Fig.~\ref{fig:Einfl_N12N768} differs from the
prediction. The modified Netz' theory that considers PEs as ellipsoidal objects
predicts that the critical electric field scales as $V^{-1/2}$ where $V$ is
calculated from the eigenvalues of the gyration tensor~\cite{hsiao08b}.  Our
refining results show that $V\sim N^{2.44(1)}$ for $C_s=0.0$ and $V\sim
N^{1.13(1)}$ for $C_s=C_s^*$ in zero fields. The modified Netz' theory hence
predicts $E^*\sim N^{-1.22(1)}$ and $E^*\sim N^{-0.56(1)}$, which are, again,
inconsistent with the results of Fig.\ref{fig:Einfl_N12N768}.  The differences
are so important that a new and detailed understanding of the mechanism of
chain unfolding is necessary.  It is definitely a topic worthy to be
investigated in the future.  

\textcolor{\RED}{We remark that Netz' original work does not verify the
scaling of $E^*$ with chain length. Moreover, the PE was collapsed by
monovalent counterions ($q=1$) due to the assumption of a large Coulomb
coupling constant $\Xi=\lambda_B q^2/\sigma$, ranging between 5 and 30. The
situation is equivalent to a collapsed PE in a (1:1)-salt solution at very low
temperature because the Bjerrum length $\lambda_B$ is large, with a value lying
between $5\sigma$ to $30\sigma$. In the derivation, no multivalent counterions
were considered, neither was the competition between multivalent ions and
monovalent ones as happens in reality. Therefore, the disagreement with our
results arises from these simplified assumptions in the model. In the theory,
the critical polarization energy is equal to $k_BT$, which has to be improved.
Moreover, the transition is assumed to be sharp.  Our simulations reveal a more
complicated story: The chain conformational transition is a continuous change
at $C_s=0.0$, while it is a more discontinuous transition at
$C_s=C_s^*$, particularly when $N$ is large.  Therefore, the onset critical
field $E^*_{II}$ and the inflection critical field $E^*_{III}$ follow different
scaling laws in different solutions. The ratio between the two estimators
describes the sharpness of the transition and theory should account for its
deviation from unity. }

\textcolor{\RED}{In our previous work~\cite{wei09}, the critical electric field
in trivalent salt solutions at the equivalence point was studied through the
inflection-point method. The scaling exponent was found to be $-0.77(1)$, which
is consistent with what we obtain here. This seems to suggest that the
inflection critical field for trivalent and for tetravalent salts follows a
similar scaling law. Liu et al.~have recently studied the scaling law of DC
unfolding fields in monovalent, divalent, and trivalent salt solutions by
simulations~\cite{liu10}. The exponents reported depended on the salt
valency. The result for trivalent salt ($-0.64$) at equivalence point is not
consistent with ours. Since the chain length is short in their study (only up
to $N=192$ in comparison with ours $N=768$) and the inflection point is
ambiguous (see Fig.~3 in the paper), their results are not reliable.  However,
the study gives rise a relevant and interesting question: Does the critical
electric field follow a different scaling law when the valency of salt is
small, such as monovalent and divalent salt, as claimed?  It necessitates a
detailed and precise investigation in the future. According to our work, the
unfolding electric field shares a similar scaling exponent in trivalent and
tetravalent salt systems but the prefactor in the scaling law is different.
Many phenomena in both of the systems occur in a similar way, such as ion
distributions and mobility changes. } 

\textcolor{\RED}{We study the tetravalent system because large ion valency can
give a stronger response to applied electric fields, which makes observing the
effects easier. From the point of view of applications, it is very important
to understand the role of salt valency in PE solutions in electric fields.  The
selection of salt valency depends on the research context and goal.  We hope
that the information reported here can be helpful in the development of
techniques for molecular separation and in the design of functionalized
micro/nano-fluidic devices.}

\section{Conclusions}
In this work, we performed molecular dynamics simulations to investigate the
conformational and electrophoretic properties of chains in tetravalent salt
solutions subject to electric fields. Our results show that chain size depends
strongly on the salt concentration $C_s$ and that under the action of electric
fields, the chain shape can be altered. When the field strength is stronger
than a critical value $E^*$, the chains are largely extended to an elongated
structure. \textcolor{\BLUE}{Two estimators $E_{I}^*$ and $E_{II}^*$ were used
to calculate $E^*$ through equating $W_{pol}$ to $k_BT$ and identifying the
onset point of $D_s$, respectively. The obtained values show that $E^*$ is a
non-monotonic function of $C_s$, and the maximum value appears at the
equivalence point $C_s=C_s^*$. The dipole moment shows that chain polarization
displays a linear dependence on the electric field up to $E^*_{II}$.}

The salt concentration has a strong influence on the electrophoretic mobility
of the chain and ion distributions. In weak electric fields, the chain mobility
$\mu_{pe}$ is negative for $C_s<C_s^*$, whereas it is positive when
$C_s>C_s^*$. The latter demonstrates the sign inversion of the effective chain
charge.  The mobility of the condensed tetravalent counterions $\mu^c_{+4}$ is
identical to $\mu_{pe}$ because the condensed ions are tightly bound onto the
chains and move together in the electric fields.  When the applied electric
field $E$ is stronger than $E_{II}^*$, the chains unfold such that $\mu^c_{+4}$ and
$\mu_{pe}$ are no longer identical.  In an even higher $E$ field, $\mu^c_{+4}$
becomes positive and $\mu_{pe}$ becomes negative: There is a relative motion
between the chains and the condensed tetravalent ions. The number of the
condensed ions shows that part of the condensed tetravalent ions are
stripped-off from the chains due to the strong electric fields. The behavior of
the effective chain charge $Q$ is consistent with the chain mobility found.

Moreover, we studied in detail the distribution of different species of ions
condensed on the chains. \textcolor{\BLUE}{The counterions are dragged toward
the rear of the chain, due to the polarization. For $C_s<C_s^*$, the
condensation profile displays that the tetravalent counterions condense at the
chain rear, followed by the condensed monovalent ones right behind the
tetravalent, close to the chain center, owing to the strong electrostatic
repulsion between the counterions.  For $C_s\ge C_s^*$, monovalent counterions
are totally expelled from the chain, which leaves only the condensed
tetravalent counterions, which overcharge the surface of the chains due to the
excessive condensation.  Consequently, the coions are attracted to the chains
and condensed. This process is mediated by the condensed tetravalent
counterions because a similar profile of coion distribution to the tetravalent
one is found. An analogy of our system to sedimentation problems was  used to
explain the ion condensation profile. } 

Finally, we investigated the dependence of chain mobility and unfolding
transition on the chain length $N$. \textcolor{\BLUE}{The estimator $E_{III}^*$
was considered through the calculation of the inflection point of
$D_s$.  At $Cs=0.0$, the transition of chain size is not sharp.  But $E_{II}^*$
and $E_{III}^*$ follow a similar scaling law $N^{-0.63(4)}$ with different
prefactors. However, at the equivalence point, the unfolding transition becomes
sharper. $E_{III}^*$ scales as $N^{-0.76(2)}$. A crossover was observed in
$E_{II}^*$, which converges asymptotically to $E_{III}^*$ when $N$ is large.
The scaling laws obtained here are significantly different to both Netz'
prediction and the modified theory.  Noticeably, $E_{I}^*$ fails to predict the
unfolding field, which shows that the critical polarization energy is not
simply $k_BT$.  It hence necessitates a further investigations to explore the
mechanism of chain unfolding in electric fields and the associated mobility and
electrokinetics changes.}

\section*{Acknowledgments}
We thank T.~N.~Shendruk for reading the manuscript and discussing.  This material
is based upon work supported by the National Science Council, the Republic of
China, under Grant No.~NSC 97-2112-M-007-007-MY3. Computing resources are
supported by the National Center for High-performance Computing.


\end{document}